\documentclass[twocolumn,superscriptaddress,amsmath, amssymb, amsfonts,preprintnumbers,aps,prd,longbibliography,nofootinbib]{revtex4-2}

\renewcommand{\sec}[1]{\noindent\textbf{#1. --- }}

\usepackage{graphicx}
\usepackage{dcolumn}
\usepackage{bm}
 \usepackage{amstext}
 \usepackage{amssymb}
 \usepackage{amsmath}
 \usepackage{graphicx}
 \usepackage{color}
 \usepackage{bbold}
 \usepackage{delimset} 
\usepackage[caption=false,justification=justified]{subfig}
\usepackage[colorlinks=true]{hyperref}
\usepackage{orcidlink}
\usepackage{float}
\usepackage[normalem]{ulem}
\usepackage{mathtools}
\usepackage[colorlinks=true]{hyperref}

\newcommand{\CricArrowRight}[1]{%
    \setlength{\@SizeOfCirc}{\maxof{\widthof{#1}}{\heightof{#1}}}%
    \tikz [x=1.0ex,y=1.0ex,line width=.15ex, draw=blue]%
        \draw [->,anchor=center]%
            node (0,0) {#1}%
            (0,1.2\@SizeOfCirc) arc (85:-240:1.2\@SizeOfCirc);%
}%

\usepackage{tikz}
\usetikzlibrary{decorations.pathmorphing}
\usetikzlibrary{decorations.markings}
\usetikzlibrary{positioning, shapes, snakes}
\usetikzlibrary{arrows.meta}

\tikzset{
	graviton/.style={line width=.8pt, -latex,decorate, decoration={snake, segment length=4pt,amplitude=1.8pt, pre length=.1cm, post length=.25cm}},
	worldline/.style={gray, line width=1pt},
	worldlineBold/.style={black, line width=.6pt},
	zUndirected/.style={line width=1pt},
	zParticle/.style={line width=1pt,postaction={decorate},decoration={markings,mark=at position .6 with {\arrow[#1]{latex}}}},
	zParticleBold/.style={line width=1.5pt,postaction={decorate},decoration={markings,mark=at position .65 with {\arrow[#1]{latex}}}},
	zParticleF/.style={line width=1pt,postaction={decorate}},
	cscalar/.style={line width=1pt,postaction={decorate},decoration={markings,mark=at position .6 with {\arrow[#1]{latex}}}},
	cscalar2/.style={line width=1pt,postaction={decorate},decoration={markings,mark=at position .8 with {\arrow[#1]{latex}}}},
	photon/.style={line width =.8pt, decorate, decoration={snake, segment length=4pt, amplitude=1.8pt,  pre length=.1cm, post length=.1cm}},
	cutZ/.style={line width=1pt,postaction={decorate},decoration={markings,mark=at position .6 with {\arrow[#1]{Rays}}}},
	invisArrow/.style={postaction={decorate},decoration={markings,mark=at position .6 with {\arrow[#1]{latex}}}},
}

\newcommand{\ii}{\text{i}}
\newcommand{\zz}{0^+}

\DeclareFontFamily{OT1}{pzc}{}
\DeclareFontShape{OT1}{pzc}{m}{it}{<-> s * [1.350] pzcmi7t}{}
\DeclareMathAlphabet{\mathpzc}{OT1}{pzc}{m}{it}

\def\cO{\mathcal{O}}

\def\eps{\epsilon}

\def\d{\mathrm{d}}

\renewcommand{\i}{\ensuremath{\mathrm{i}}}

\def\dd{\delta\!\!\!{}^-\!}

\def\d{\mathrm{d}}
\def\eps{\epsilon}

\def\nn{\nonumber}

\def\eqn#1{Eq.~\eqref{#1}}

\newcommand*\Bell{\ensuremath{\boldsymbol\ell}}

\newcommand{\widebar}{\overline}

\newcommand{\be}{\begin{equation}}
\newcommand{\ee}{\end{equation}}
\newcommand{\ba}{\begin{align}}
\newcommand{\ea}{\end{align}}
\ifx\genfrac\sdflkaj\else\fi
\newcommand{\sfrac}[2]{{\textstyle\frac{#1}{#2}}}

\newcommand{\gam}{\gamma}

\newcommand{\Gam}{\Gamma}

\newcommand{\varK}{\varpi_{0,\rm K3}}
\newcommand{\varCY}{\varpi_{0}}

\begin{document}

\preprint{HU-EP-24/32-RTG}
\preprint{QMUL-PH-24-26}
\preprint{BONN-TH-2024-15}
\preprint{TUM-HEP-1532/24}

\title{Emergence of Calabi-Yau manifolds in high-precision black hole scattering}
\author{Mathias Driesse\,\orcidlink{0000-0002-3983-5852}} 
\affiliation{%
Institut f\"ur Physik, Humboldt-Universit\"at zu Berlin,
10099 Berlin, Germany
}

\author{Gustav Uhre Jakobsen\,\orcidlink{0000-0001-9743-0442}} 
\affiliation{%
Institut f\"ur Physik, Humboldt-Universit\"at zu Berlin,
10099 Berlin, Germany
}
\affiliation{Max Planck Institut f\"ur Gravitationsphysik (Albert Einstein Institut), 14476 Potsdam, Germany}

\author{Albrecht Klemm\,\orcidlink{0000-0001-5499-458X}} 
\affiliation{%
Bethe Center for Theoretical Physics, Universit\"at Bonn, 53115 Bonn, Germany
}
\affiliation{%
Hausdorff Center for Mathematics, Universit\"at Bonn, 53115 Bonn, Germany
}

\author{Gustav Mogull\,\orcidlink{0000-0003-3070-5717}}
\affiliation{%
Institut f\"ur Physik, Humboldt-Universit\"at zu Berlin,
10099 Berlin, Germany
}
\affiliation{Max Planck Institut f\"ur Gravitationsphysik (Albert Einstein Institut), 14476 Potsdam, Germany}
\affiliation{Centre for Theoretical Physics, Department of Physics and Astronomy, Queen Mary University of London,  London E1~4NS, United Kingdom}

\author{Christoph Nega\,\orcidlink{0000-0003-0202-536X}}
\affiliation{%
Physik Department, Technische Universit\"at M\"unchen, 85748 Garching, Germany
}
  
 \author{Jan Plefka\,\orcidlink{0000-0003-2883-7825}} 
\affiliation{%
Institut f\"ur Physik, Humboldt-Universit\"at zu Berlin,
10099 Berlin, Germany
}

\author{Benjamin Sauer\,\orcidlink{0000-0002-2071-257X}} 
\affiliation{%
Institut f\"ur Physik, Humboldt-Universit\"at zu Berlin,
10099 Berlin, Germany
}

\author{Johann Usovitsch\,\orcidlink{0000-0002-3542-2786}} 
\affiliation{%
Institut f\"ur Physik, Humboldt-Universit\"at zu Berlin,
10099 Berlin, Germany
}

\begin{abstract}
Using the worldline quantum field theory formalism,
we compute the radiation-reacted impulse, scattering angle, radiated energy and
recoil of a classical black hole (or neutron star) scattering event at fifth
post-Minkowskian and sub-leading self-force orders (5PM-1SF).
This state-of-the-art four-loop computation employs advanced integration-by-parts
and differential equation technology, and is considerably more challenging
than the conservative 5PM-1SF counterpart.
As compared with the conservative 5PM-1SF,
in the radiation sector Calabi-Yau three-fold periods appear and
contribute to the radiated energy and recoil observables.
We give an extensive exposition of the canonicalization of the differential equations
and provide details on boundary integrations, Feynman rules, and integration-by-parts strategies.
Comparisons to numerical relativity are also performed.
\end{abstract}
 
\maketitle 


When two massive objects (black holes, neutron stars or stars) in our universe fly past each
other, their gravitational interactions deflect their trajectories~\cite{Einstein:1918btx,Kovacs:1978eu}. The gravitational waves emitted in the related bound-orbit system — the binary inspiral — are now routinely detected by gravitational wave observatories~\cite{KAGRA:2021vkt}. Theoretical physics needs to provide high-precision templates to make use of unprecedented sensitivity and precision of the data from upcoming gravitational wave observatories~\cite{Purrer:2019jcp}. Motivated by this challenge, a number of analytical and numerical techniques have been developed to approximately solve this gravitational two-body problem. While numerical relativity is accurate~\cite{Pretorius:2005gq,Damour:2014afa,Boyle:2019kee}, it is too time-consuming to rapidly produce large numbers of gravitational wave templates. For this, approximate analytical results are also required~\cite{Blanchet:2013haa,Porto:2016pyg,Levi:2018nxp,Mogull:2020sak,Kosower:2022yvp,Bjerrum-Bohr:2022blt,Buonanno:2022pgc,DiVecchia:2023frv}. Here we report on a new highest-precision analytical result for the scattering angle, radiated energy, and recoil of a black hole or neutron star scattering encounter at the fifth order in Newton’s gravitational coupling G, assuming a hierarchy in the two masses. This is achieved by modifying state-of-the-art techniques for the scattering of elementary particles in colliders to this classical physics problem in our universe. Our results show that mathematical functions related to Calabi-Yau manifolds, $2n$-dimensional generalizations of tori, appear in the solution to the radiated energy in these scatterings. We anticipate that our analytical results will allow the development of a new generation of gravitational wave models, for which the transition to the bound-state problem through analytic continuation and strong-field resummation will need to be performed.

\section*{Main}

Shortly after writing down his theory of general relativity, Einstein postulated the existence of gravitational waves~\cite{Einstein:1918btx}.
Just as accelerated charges give rise to
electromagnetic waves, so do accelerated masses generate gravitational radiation.
It took one hundred years to reach the 
technical ability to first detect gravitational waves \cite{LIGOScientific:2016aoc} as they emerge from binary inspirals and mergers of black holes and neutron stars  ---
the highest mass density objects in our universe.
Today, over a hundred such events have been observed by the LIGO, Virgo and KAGRA detectors \cite{KAGRA:2021vkt}.
The planned third generation of ground- and space-based gravitational wave observatories \cite{LISA:2017pwj,Punturo:2010zz,Ballmer:2022uxx},
including the recently approved LISA mission,
will reach an experimental accuracy enabling unprecedented insights into gravitational, astrophysical, nuclear, and fundamental physics.

To benefit from the increased sensitivity of gravitational wave detectors,
corresponding increases in precision are required in our
ability to solve the gravitational two-body problem~\cite{Purrer:2019jcp},
described by the highly non-linear Einstein field equations,
and thus predict the gravitational waves produced in a binary encounter.
While numerical relativity, which discretizes spacetime and solves the resulting equations numerically on supercomputers, 
provides a good option~\cite{Pretorius:2005gq,Damour:2014afa,Boyle:2019kee}, it is slow and computationally expensive (a run for a single configuration can take weeks).
As tens of millions of waveform templates are needed for gravitational wave data analysis,
fast approximate analytical results to the two-body problem are therefore also required.
These can be generated using perturbation theory,
picking one or more small parameters and solving the equations order-by-order in a series expansion.
For the binary inspiral this includes a slow-velocity and weak gravitational field
expansion (post-Newtonian) \cite{Blanchet:2013haa,Porto:2016pyg,Levi:2018nxp} as well as the semi-analytical self-force expansion \cite{Poisson:2011nh,Barack:2018yvs,Gralla:2021qaf} in a small mass ratio ($m_{1}/m_{2}\ll 1$).

\begin{figure}[t] 
	\includegraphics[width=0.8\hsize]{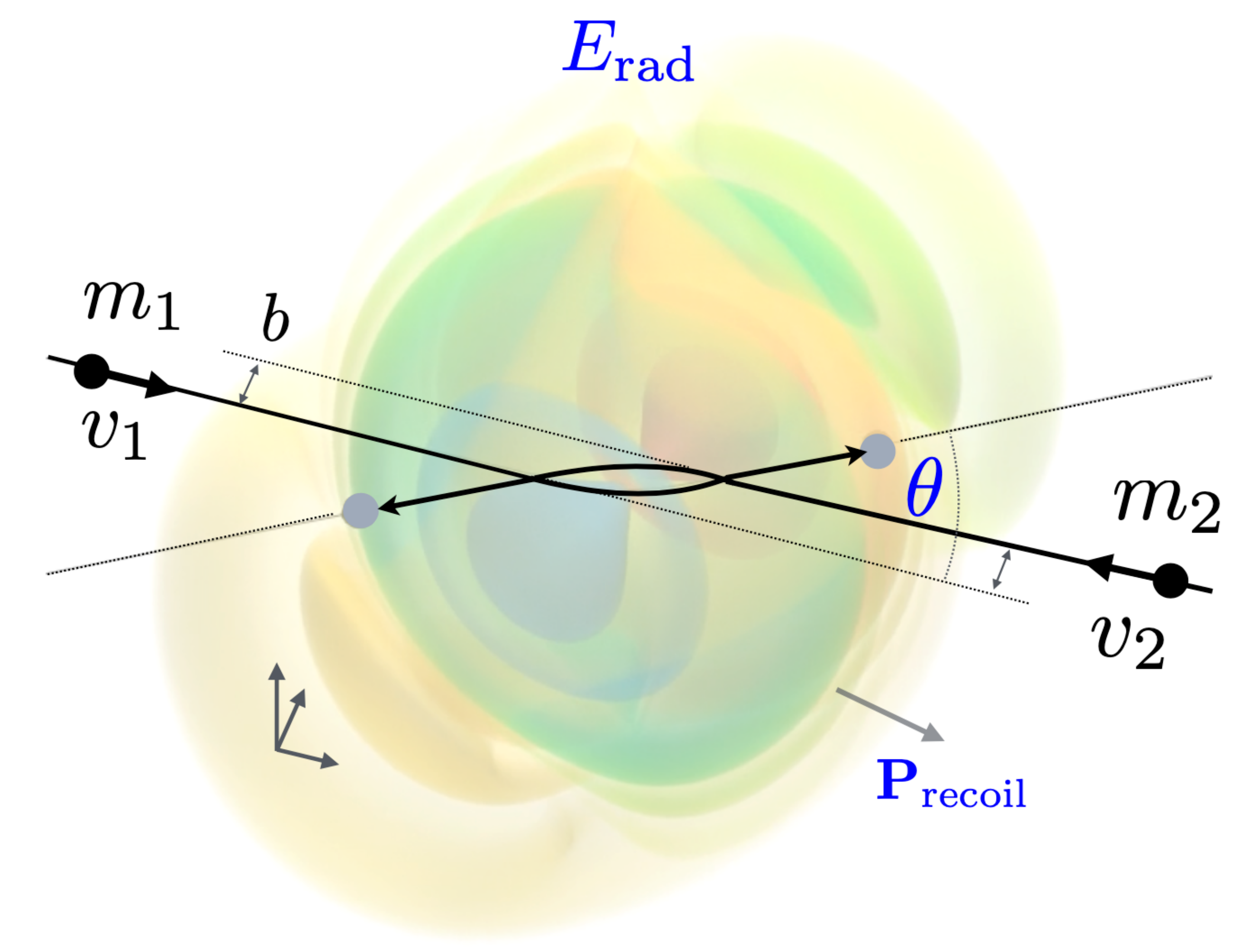}
	 \caption{{\bf Gravitational two-body scattering event:} Two black holes (or neutron stars) with masses $m_{i}$ and incoming velocities $v_{i}$, impact parameter $b$ and resulting
 relative scattering angle $\theta$, radiated gravitational wave energy $E_{\text{rad}}$ and recoil.}
	  \label{fig-1}
\end{figure}

Here we describe a black hole (or neutron star) scattering encounter,
which --- while asymptotically unbounded ---
also yields physical data concerning the bound two-body problem, relevant for binary inspirals~\cite{Kalin:2019rwq,Buonanno:2024byg}.
In the scattering regime one may advantageously adopt a weak gravitational field expansion, in powers of Newton's constant $G$, 
valid as long as the two bodies are well
separated but moving at arbitrary velocities~\cite{Kalin:2020mvi,Mogull:2020sak,Kosower:2022yvp,Bjerrum-Bohr:2022blt,Buonanno:2022pgc,DiVecchia:2023frv} --- see Fig.~\ref{fig-1}. 
The first non-trivial order for such a black hole scattering was found in 1979 \cite{Westpfahl:1985tsl}, namely the 
sub-leading $G^2$ order \cite{Damour:2017zjx}.
Rapid progress has been made since then by synergistically
porting techniques from quantum field theory (QFT)~\cite{Goldberger:2004jt,Bjerrum-Bohr:2022blt,Kosower:2022yvp,Buonanno:2022pgc},
the mathematical framework for elementary particle scattering,
to this classical physics problem.
The third order ($G^{3}$) was established in 2019 \cite{Bern:2019nnu,Damour:2019lcq,Kalin:2020fhe,DiVecchia:2021bdo}
and the fourth order ($G^{4}$) was completed in 2023 \cite{Bern:2021yeh,Dlapa:2021vgp,Dlapa:2022lmu,Damgaard:2023ttc,Jakobsen:2023hig}.
At least the fifth order ($G^{5}$) precision will be needed in order to prepare for the third generation of gravitational wave detectors~\cite{Purrer:2019jcp}.

Being characterized by three fundamental properties --- their mass, spin and charge  ---
black holes are, in a sense, the astrophysical equivalents of elementary particles.
QFT is a highly mature subject,
and precise analytic predictions for particle scattering events,
used at colliders, such as CERN's Large Hadron Collider, are commonplace.
We benefit from this progress in gravity through the close theoretical link between hyperbolic motion (unbound scattering)
and elliptic motion (bound orbits).
State-of-the-art technologies for performing the multi-loop Feynman integrals involved in  scattering cross sections,
have enabled some remarkable predictions in elementary particle physics~\cite{Anastasiou:2015vya,Jones:2018hbb,Chen:2021isd,Lee:2022nhh} and uncovered surprising connections to algebraic 
geometry~\cite{Bogner:2007mn,Vanhove:2014wqa,Bourjaily:2018yfy,Bonisch:2021yfw}.

\begin{figure}[t] 
	\includegraphics[width=1.0\hsize]{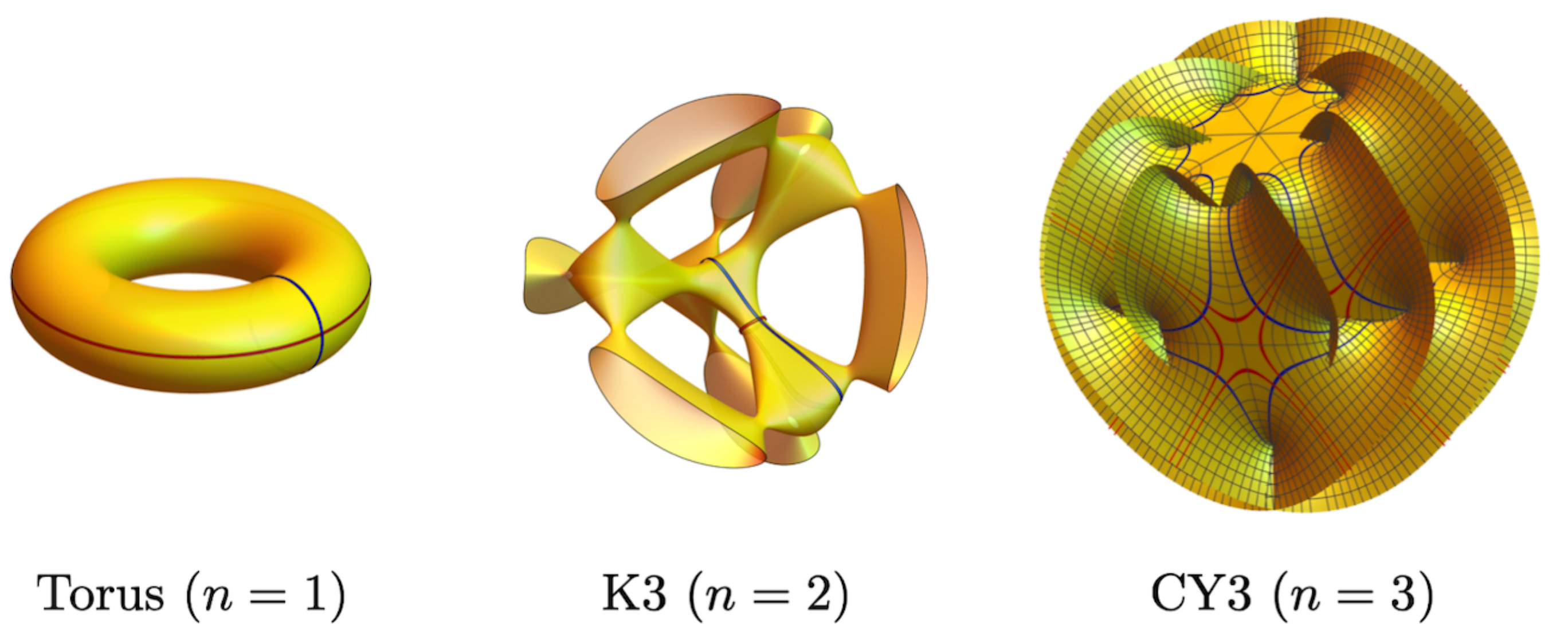}
	 \caption{	{\bf  Graphical representation of the Calabi-Yau (CY) $n$-folds emerging in
	the black hole scattering:}  The elliptic curve (topologically a torus), and two-dimensional projections of the K3 surface and CY3 reflecting their symmetries. Red and blue lines are (projections of) the real $n$-dimensional cycles $\Gamma_n$.
	The corresponding periods over the $n$-form $\Omega_n(x)$, i.e. $\int_{\Gamma_n} \Omega_n(x)$, depend on the so-called modulus $x$ (related to the relative velocity of the black holes
	 $v_{1}\cdot v_{2}/c^{2}=(x+x^{-1})/2)$ parametrizing the shape of CYs and yield master integrals in our problem.	}
	  \label{fig-5}
	  \end{figure}

The link to algebraic geometry arises through the function spaces that are needed to express the observables at growing perturbative orders. 
One typically finds generalizations of logarithms, known as multiple polylogarithms,
that are well understood. At higher orders, elliptic integrals make their appearance~\cite{Bourjaily:2022bwx}. 
Geometrically, these are period integrals over the two non-trivial closed cycles of a family of elliptic curves (also known as tori)--- see Fig.~\ref{fig-5}. 
The physical parameters determine the shape of the latter. Yet this is just the tip of the iceberg. 
 
Recently, it has become evident~\cite{Vanhove:2014wqa, Bonisch:2020qmm, Bonisch:2021yfw, Duhr:2022pch, Bourjaily:2018yfy} that Calabi-Yau (CY) manifolds emerge in generalizations of the aforementioned function spaces, which encode Feynman integrals in higher-order perturbation theory. 
These are complex $n$-dimensional manifolds whose metric obeys Einstein’s vacuum equations in $2n$-spacetime dimensions~\cite{MR480350}.
Geometrically, these higher-dimensional CY $n$-folds represent a beautiful series of critical geometries
generalizing the elliptic curve ($n=1$), and may be thought of as $2n$-dimensional generalizations of the torus. To motivate this, consider the Legendre family of elliptic curves $Y^2=X(X-1)(X-x)$ with $X,Y$ complex variables. The one-form $\Omega_1= {\rm d}X/Y$ yields
the elliptic periods $\varpi_0=2\int_1^\infty{\rm d}X/Y$ and $\varpi_1=-2i\int_{-\infty}^0{\rm d}X/Y$ which are expressible through standard elliptic integrals. 
These satisfy a second-order differential equation $\left( 1+4(2x-1)\partial_x+4x(x-1)\partial_x^2 \right)\varpi_k=0$ for $k=0,1$, known as a Picard-Fuchs equation.
In turn, CY $n$-folds exhibit an $n$-form $\Omega_n$, whose periods --- integrals over higher dimensional integration cycles (see Fig.~\ref{fig-5}) --- generalize the elliptic integrals. These $n$-fold periods obey Picard-Fuchs equations of order $(n+1)$.
While CY two-folds --- known as K3 surfaces~\cite{MR3586372} --- have a unique topology,
the topological types of CY ($n > 2$)-folds are not classified but believed to be finite.
CY three-folds (CY3) are of particular interest in string theory, where they are used to curl
up the six-extra spacetime dimensions to arrive at the
four observable spacetime dimensions~\cite{Candelas:1985en}.

While specific higher-loop Feynman integrals are known to be expressed in terms of CY periods~\cite{Vanhove:2014wqa, Bonisch:2020qmm, Bonisch:2021yfw, Duhr:2022pch, Bourjaily:2018yfy}, physical observables tend to be much simpler than the multitude of contributing Feynman integrals. 
For example, the Feynman integrals occurring in black hole scattering at order $G^4$~\cite{Bern:2021yeh, Dlapa:2021vgp} and $G^5$\cite{Klemm:2024wtd} encode K3 periods, but these contributions cancel spectacularly in the physical observable within the conservative sector at $G^5$~\cite{Driesse:2024xad}. 
Similar intriguing cancellations occur in quantum field theory computations~\cite{Gehrmann:2024tds, Duhr:2024bzt}. 
Furthermore, CY $n$-fold periods have a transcendentality degree~\cite{MR2954618, Bonisch:2021yfw} increasing with their dimension $n$. 
This leads to the important question of what classes of transcendental functions appear in physical observables in perturbation theory. 
Before our work, no physical observables were known that feature CY $n$-fold periods for $n\ge 3$. 
One expects that our findings and the methods described below will have significant implications for high-precision predictions in particle physics as well.   

In this article, we report on a new landmark result of the QFT-based classical general relativity
program by providing complete scattering observables 
of a binary black hole (or neutron star) encounter up to fifth order in the weak-field expansion ($G^5$),
and sub-leading order in the symmetric mass ratio $\nu=m_1m_2/(m_1+m_2)^2$.
This encounter is depicted in Fig.~\ref{fig-1},
and involves two black holes scattering with a deflection angle $\theta$ and radiating gravitational waves with total energy $E_{\rm rad}$.
We describe the black holes as point particles,
an approximation valid as long as their separation $b$ is large compared to their intrinsic sizes, i.e.~their Schwarzschild radii $2Gm_i/c^2$ ---
the weak gravitational field region. 
Consequently, the $G$ expansion is really an expansion in the dimensionless quantity
$GM/bc^{2}$ with total mass $M=m_{1}+m_{2}$.  
The two scattering observables $\theta$ and $E_{\rm rad}$,
the latter depending on CY3 periods, can be used to calibrate gravitational waveform models.

The gravitationally interacting two-body system
is governed by an action comprised of two worldlines, coupled to the gravitational Einstein-Hilbert term:
\begin{align}\label{eq:action}
S=-m_1c\!\int\!\d s_1-m_2c\!\int\!\d s_2
- \frac{c^3}{16\pi G} \int\!\d^4x \sqrt{-g}\, R[g] \,.
\end{align}
Variation of this action gives rise to the Einstein and geodesic equations.
To explain our notation: the proper time intervals $\d s_i=\sqrt{g_{\mu\nu}\dot x_i^{\mu}\dot x_i^{\nu}}\d\tau$ give rise to the followed trajectories $x_{i}^{\mu}(\tau)$ ($\mu=0,1,2,3$) of the $i$'th black hole,
parametrized by a time parameter $\tau$ (a dot symbolizes a $\tau$ derivative). The spacetime metric $g_{\mu\nu}(x)$ yields the curvature scalar $R[g]$ and $g=\det(g_{\mu\nu})$.

We calculate the change in four-momentum of each body over the course of scattering, $\Delta p_i^\mu$,
known as the impulse.
With the momentum of each body given by $p_i^\mu=m_i\dot{x}_i^\mu$,
the impulse is simply the difference between the momentum at late and early times:
\begin{align}\label{eq:impExp}
	\Delta p_i^\mu&=p_i^\mu(\tau\to+\infty)-p_i^\mu(\tau\to-\infty)\\
	&=G\Delta p_i^{(1)\mu}+G^2\Delta p_i^{(2)\mu}+\cdots+G^5\Delta p_i^{(5)\mu}+\cdots\,.\nn
\end{align}
The initial momentum of each black hole is given by its mass times initial velocity: $p_i^\mu(\tau\to-\infty)=m_iv_i^\mu$.
Working in a weak gravitational field region
we have series expanded order-by-order in Newton's constant.
With results up to $G^4$ already determined~\cite{Bern:2021yeh,Dlapa:2022lmu,Damgaard:2023ttc},
and the conservative (non-radiating) part of $G^5$ derived by some of the current authors~\cite{Driesse:2024xad},
here we extract the subleading-in-$\nu$ $G^5$ component
from which we will also derive the scattering angle $\theta$ and radiated energy flux $E_{\rm rad}$.
Note that $\nu$ tends to zero for $m_{1}\ll m_{2}$ and vice versa.

Our calculation is performed using  Worldline QFT (WQFT)~\cite{Mogull:2020sak,Jakobsen:2022psy},
wherein a Worldline Effective Field Theory action is used to represent the black holes as point particles~\cite{Goldberger:2004jt,Kalin:2020mvi}.
This allows us to re-interpret this classical physics problem
as one of drawing and calculating perturbative Feynman diagrams ---
see Fig.~\ref{fig-2}.
The WQFT's main benefit for classical physics computations
is a clean separation between classical and quantum effects.
In this language, gravitons (wavy lines) and deflection modes (solid lines)
are the quantized excitations of the metric $g_{\mu\nu}$ and trajectories $x_i^\mu$.
These particles' momenta and energies are unfixed, and must be integrated over.
The key principle being exploited here is that tree-level one-point functions,
 given by a sum of diagrams with a single outgoing line and no internally closed loops,
 solve the classical equations of motion \cite{Boulware:1968zz}.
 We recursively generated the graphs to be computed at the fifth order in $G$,
yielding a total of 426 diagrams.
These diagrams directly translate to mathematical expressions, Feynman integrals,
by way of Feynman rules derived from the action~\eqref{eq:action} --- see Fig.~\ref{fig-3}.

The resulting Feynman integrals are a staple of perturbative QFT. 
Individual Feynman integrals, which may diverge in four spacetime dimensions,
are treated by working in $D$ dimensions so that divergences appear as $(D-4)^{-1}$ poles.
Finiteness of our results in the limit $D\to4$, i.e.~the cancellation of all intermediate divergences,
then provides a useful consistency check.
Our calculation of the impulse calls for the evaluation of millions of Feynman integrals,
which may have at most 13 propagators of the kinds seen in Fig.~\ref{fig-3}.
To evaluate them we generate linear integration-by-parts (IBP) identities 
which reduces the problem to one solving a large system of linear equations.
The task was nevertheless enormous,
and consumed around 300,000 core hours on HPC clusters.

\begin{figure}[t] 
	\includegraphics[width=1.0\hsize]{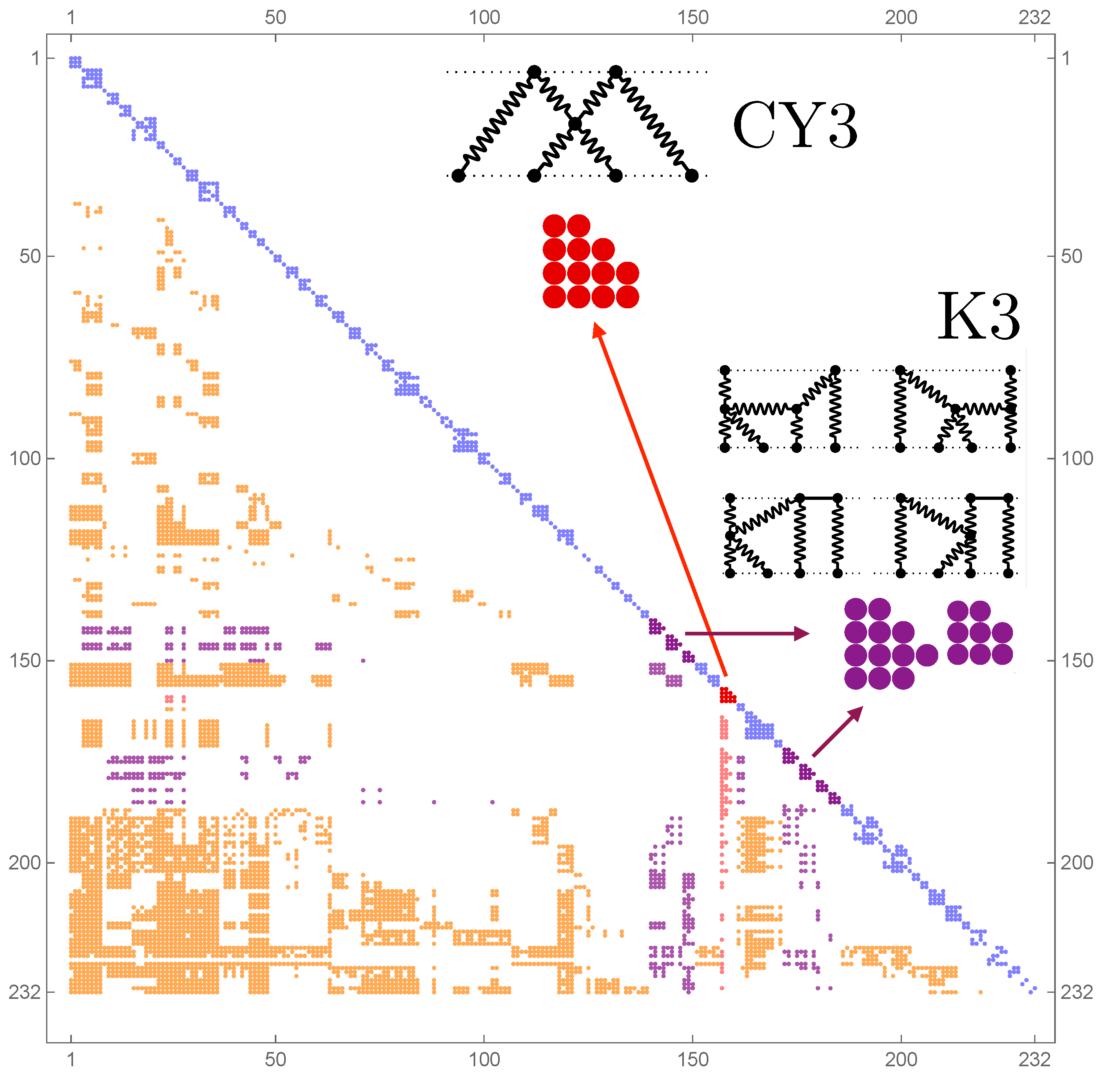}
	 \caption{{\bf Non-zero entries of the odd parity $232\times 232$ differential equation matrix 
$\mathbf{\hat{M}(x, D)}$:} The blocks on the diagonals
	 determine the function spaces of the multiple sub-sectors. The not-magnified diagonal sectors give rise to multiple polylogarithms.
	}
	  \label{fig-4}
	  \end{figure}

Our task is ultimately to determine expressions for a basis of 236+232 master integrals,
that split under parity ($v_{i}^{\mu}\to -v_{i}^{\mu}$) into two distinct sectors. From these, all other integrals may be expressed as linear combinations using the IBP identities.
To do so, we exploit the integrals' non-trivial dependence on only a single variable $x$: 
it derives from 
the relativistic boost factor $\gamma=1/\sqrt{1-(v/c)^{2}}=v_{1}\cdot v_{2}/c^{2}$ for the initial relative velocity $v$ of the two black holes,
via $\gamma = (x+x^{-1})/2$. 
Rather than attempt a direct integration,
we may therefore set up two systems of differential equations in $x$ (even and odd parity) as 
\begin{align}\label{eq:diffEqn}
\frac{\d}{\d x} \vec{I}(x,D) = \hat{M}(x, D)\, \vec{I}(x,D)\,.
\end{align}
The integrals to be computed are grouped into vectors $\vec{I}$,
and the matrices $\hat{M}$ take a lower block-triangular form --- see Fig.~\ref{fig-4}.
To obtain this system, derivatives of the master integrals with respect to $x$ are re-expressed as linear combinations
of the masters themselves using the IBP identities.
We solve the system order-by-order in a series expansion close to $D=4$,
with higher-order terms given by repeated integrals (with respect to $x$)  of lower-order terms.
Boundary conditions on the integrals are fixed in the non-relativistic (low-velocity) limit $x\to 1$.

Repeated integrations with respect to the kinematic parameter $x$ produce the
mathematical functions ${\cal I}$ appearing in our final result for the impulse:
\begin{align}\label{eq:integralDef}
	{\cal I}(\phi_1,\phi_2,\ldots,\phi_n;x)
	\coloneqq\!\!
	 \int_1^x\!\!
	 \d x'
	 \phi_1(x'){\cal I}(\phi_2,\ldots,\phi_n;x'),
\end{align}
with the base case ${\cal I}(;x)=1$. The nature of the integration kernels $\phi_i$ determines the types of functions,
and are associated with underlying geometries. In the simplest case, the kernels $\phi_i$ are rational functions with single poles, e.g., $x^{-1}$, $(x+1)^{-1}$ or $(x-1)^{-1}$, and iterated integrations produce the function class of multiple polylogarithms ---
including the ordinary logarithm ${\cal I}(x^{-1};x)=\log x$. Geometrically, one can interpret these integration kernels as periods of a zero-dimensional CY space,
given by two points on a sphere.
In more complicated scenarios, usually related to higher-loop computations, the $\phi_i$ are connected to periods of higher-dimensional algebraic varieties. A key challenge is to understand the 
kernels and the associated class of iterated integrals occurring in a physical problem.  
In a $G^4$ calculation of the impulse, squares of elliptic integrals arise, which are geometrically interpreted as periods of a one-parameter K3 surface --- see Fig.~\ref{fig-5}.
In the odd parity sector of integrals at the present fifth order in $G$, the kernels also depend
on CY3 periods and we express physical quantities in terms of the corresponding class of iterated functions.

To see the origin of the CY3 periods, and to clarify their precise nature,
we examine the differential equation matrix $\hat{M}(x,D)$ (see Fig.~\ref{fig-4}) in the limit $D\to4$.
The diagonal blocks of this matrix are associated with specific Feynman graphs appearing in the impulse,
of which the CY3 geometry is isolated to a single $4\times 4$ diagonal sub-block.
One can decouple these four first-order differential equations such that one obtains a single fourth-order differential equation, which is the Picard-Fuchs equation of the CY three-fold:
\be \label{CY3}
\bigg[ \Big(x\frac{\d}{\d x} -1\Big)^4 - x^4\Big(x\frac{\d}{\d x}+1\Big)^4\bigg]\varpi(x)	=	0 \, .
\ee
The latter is solved by the four CY three-fold\footnote{The corresponding one-parameter family of CY three-folds is of hypergeometric type and has intriguing arithmetic properties discussed in ref.~\cite{MR4748808}. For an algebraic definition of the CY family together with an explicit expression of $\Omega_3$, we refer to refs.~\cite{MR4748808, Klemm:2024wtd}. It is intriguing that the occurring K3 and CY3 periods are related to those of the  Legendre curve by a symmetric and a Hadamard product, respectively~\cite{MR4748808, Klemm:2024wtd}.} periods $\varpi_k(x)=\int_{\Gamma^k_3} \Omega_3(x)$, where the three-form $\Omega_3(x)$ is integrated over the real three-dimensional cycles $\Gamma^k_3$, $k=0,1,2,3$.
These integrals appear within the integration kernels $\phi_i$ of the iterated integrals~\eqref{eq:integralDef}.

\begin{figure}[t] 
	\includegraphics[width=0.99\hsize]{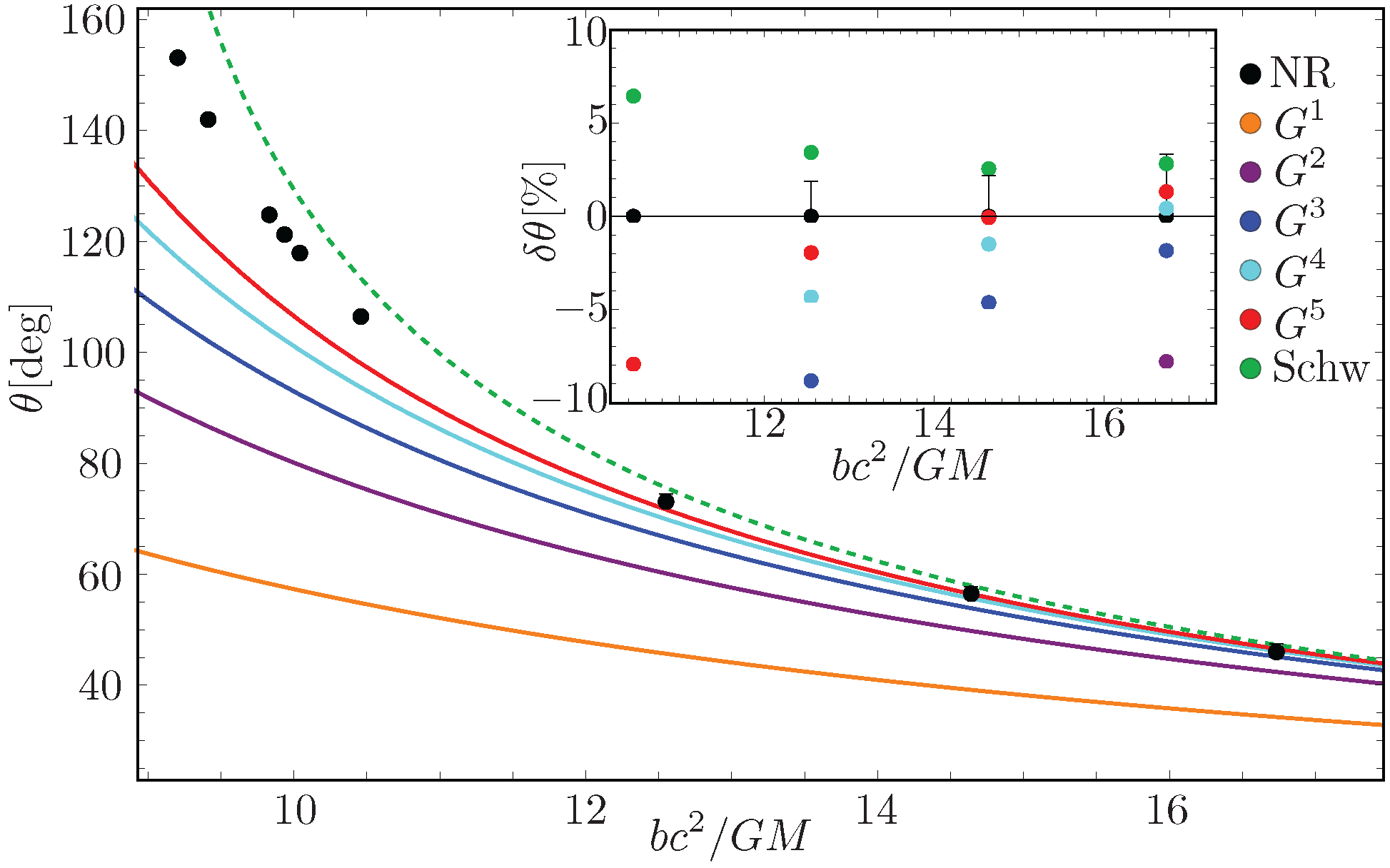}
	 \caption{{\bf The scattering angle $\mathbf{\theta}$:} Plotted as a function of the impact parameter in units of the Schwarzschild radius,
	 $bc^{2}/GM$, up to order $G^{5}$ for an equal-mass scenario with initial relative velocity $v=0.5125c$.
	    The black dots are existing numerical relativity (NR) simulations~\cite{Rettegno:2023ghr}. The $G^5$ curve follows from
	    (\ref{theta5}) (excluding the unknown $\nu^2 \theta^{(5,2)}$ contribution). The dotted line is the exact in $G$ ($\nu=0$)
	    \emph{probe} limit result for geodesic motion in a Schwarzschild background.
	    The inset plot depicts the relative differences to the numerical relativity data.
	    Larger values of $bc^{2}/GM$ correspond to the perturbative regime.
	    We find agreement with NR within the error for $bc^{2}/GM>12.5$. 
	     The monotonically falling corrections to the consecutive $G^{n}$
	 orders yield an intrinsic error estimate of our $G^{5}$ results: they are more precise than the NR data for $bc^{2}/GM>14$.}
	  \label{fig-6}
	  \end{figure}

\begin{figure*}[ht!] 
\includegraphics[width=1.0\hsize]{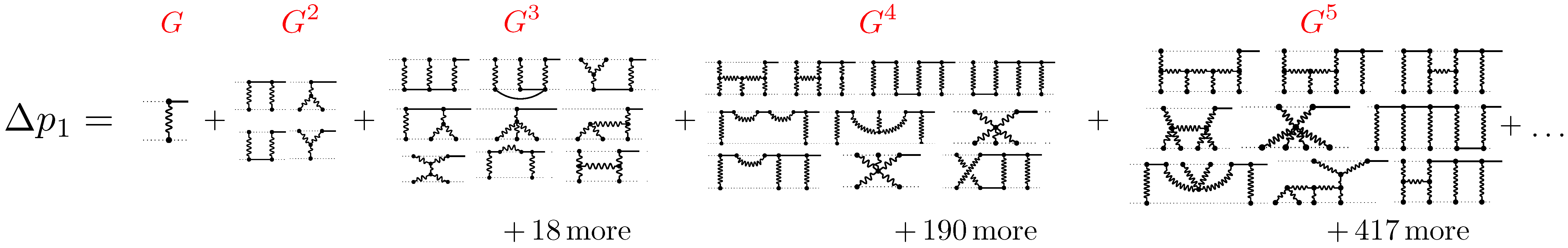}
 \caption{ {\bf Expansion of the impulse $\Delta p_{1}$ in terms of Feynman diagrams:}
 Organized order-by-order in Newton's constant $G$.
 The dotted lines represent the worldlines of the two black holes,
 exchanging gravitons (wavy lines) and propagating deflection modes (solid lines). 
}
  \label{fig-2}
\end{figure*}

\begin{figure*}[th]
	\includegraphics[width=0.8\hsize]{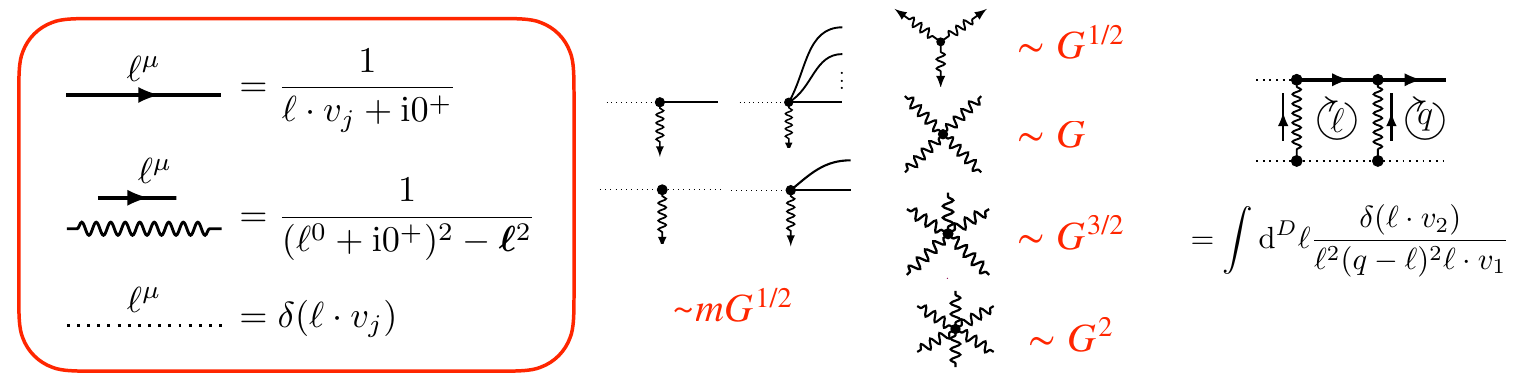}
	 \caption{{\bf Feynman rules:} Retarded graviton, retarded worldline and cut worldline propagators, the relevant worldline and bulk vertices	 and  a sample $G^2$ Feynman diagram. Here $\ell$ is the loop momentum and $q$ the momentum transfer.
}
	   \label{fig-3}
	\end{figure*}

Our final expression for the fifth-order impulse is involved and described in the Methods section,
where we also elaborate on the function space.
From the impulse we can derive the scattering angle $\theta$,
which measures the angle of deflection between the ingoing and outgoing momenta
in the initial centre-of-mass inertial frame --- see Fig.~\ref{fig-1}.\footnote{As the system dissipates energy, it recoils, and so the initial frame choice is not preserved over the course of a scattering event.}
Like the impulse~\eqref{eq:impExp}, the scattering angle is expanded in the weak-field limit
with the $G^n$ component denoted $\theta^{(n)}$.
These components are also expanded in powers of the symmetric mass ratio $\nu$, and at order $G^5$ we have
\begin{align}\label{theta5}
\theta^{(5)}=\frac{M^{5}\Gamma}{b^{5}c^{10}} \left (
\theta^{(5,0)} + \nu \theta^{(5,1)} + \nu^{2}\theta^{(5,2)} + \nu^{3} \Gamma^{-2}\theta^{(5,3)}
\right ),
\end{align}
with $M$ and $\Gamma=E/M$ the total mass and mass-rescaled energy of the initial system.
A central result of our work is the computation of all contributions except for $\theta^{(5,2)}$.
The function space of the angle $\theta^{(5)}$ arises from integrals only in the even-parity sector,
and is simpler than that of the complete impulse. We compare our result with available numerical relativity simulations \cite{Rettegno:2023ghr}
in Fig.~\ref{fig-6}.

Our other main result is the total radiated energy and momentum from the system over the course of the scattering.
Using the principle of four-dimensional momentum conservation,
which includes conservation of energy, the total loss of momentum
via gravitational wave emission must balance
the change in momenta of the two individual black holes (or neutron stars):
\begin{align}\label{Prad}
	P^\mu_{\rm rad}=-\Delta p_1^\mu-\Delta p_2^\mu\,.
\end{align}
The impulse of the second black hole, $\Delta p_2^\mu$,
can straightforwardly be inferred from that of the first using symmetry.
The radiated energy, then, is given simply by the zeroth component of the radiated four-momentum in the centre-of-mass frame:
$E_{\rm rad}=P^0_{\rm rad}=-\Delta p_1^0-\Delta p_2^0$, while the recoil ${\bf P}_{\text{recoil}}$ derives from its spatial components. 
Unlike the scattering angle, it includes integrals from the odd-parity sector,
and so contains Calabi-Yau periods. 
These terms contribute to the repeated backscattering of radiative gravitons off
the potential background -- known as the ``tail-of-tail'' effect.  

Summarizing, in this work we have extended the state-of-the-art of the gravitational two-body problem 
to a new perturbative order ($G^{5}$) to the sub-leading mass ratio level $\nu$. Our analytical findings require
the use of a new class of functions, Calabi-Yau three-fold periods, in the radiative sector. 
These methodological advances will also benefit particle phenomenology where Calabi-Yau periods appear in  higher-loop order diagrams \cite{Vanhove:2014wqa,Bonisch:2020qmm,Bonisch:2021yfw,Duhr:2022pch,Bourjaily:2018yfy}.
By comparing to
numerical relativity data we demonstrated percent-level agreement in the perturbative domain.
These results provide input data for high-precision waveform models employing effective-one-body resummation
techniques~\cite{Damour:2022ybd,Rettegno:2023ghr,Buonanno:2024vkx,Buonanno:2024byg}
that can now be developed.
For the comparable-mass case, we envision the need to also incorporate next-to-next-to-leading-order mass ratio ($\nu^{2}$) 
contributions, where novel Calabi-Yau three-folds are expected
to make their appearance \cite{Klemm:2024wtd}. This we leave for future studies.


\section*{Methods}

\sec{Integrand generation and integral family}
We employ the Worldline Quantum Field theory (WQFT) formalism \cite{Mogull:2020sak,Jakobsen:2021zvh,Jakobsen:2022psy} 
that quantizes the worldline deflections $z_{i}^{\mu}(\tau)$ and graviton field $h_{\mu\nu}(x)$ 
arising in the background field expansions $x_{i}^{\mu}(\tau) = b_{i}^{\mu}
+ v_{i}^{\mu}\tau + z_{i}^{\mu}(\tau)$ and $g_{\mu\nu}=\eta_{\mu\nu}+ \sqrt{32 \pi G} h_{\mu\nu}$, respectively (now setting $c=1$).
In the gravitational sector, we employ a non-linearly extended de Donder gauge that
simplifies the three- and four-graviton vertices (see Supplementary Information). 
The worldline actions~\eqref{eq:action} are improved by adopting the proper time gauge $\dot x_{i}^{2}=1$,
for the $i$th black hole (BH):
\be
\label{wqftact}
S_{i} = -\frac{m_{i}}2\int\!\mathrm d\tau \,g_{\mu\nu}[x_{i}(\tau)]\, \dot x_{i}^{\mu}(\tau)\dot x_{i}^{\nu}(\tau)\,.
\ee
This ensures a linear coupling to the graviton $h_{\mu\nu}$.
At the current four-loop ($G^{5}$ or 5PM) order we require up to six-graviton vertices
that derive from the Einstein-Hilbert action plus gauge-fixing term ---
taken in $D=4-2\eps$ dimensions.
We also require the single-graviton emission plus $(0,\ldots, 5)$-deflection vertices derived from Eq.~\eqref{wqftact}.
We provide the explicit vertices and graviton gauge-fixing function in a
\href{https://doi.org/10.5281/zenodo.14604438}{\tt zenodo.org} repository submission \cite{zenodo}
accompanying this article;
an analytic expression for the $n$-deflection worldline vertex was given in refs.~\cite{Mogull:2020sak,Jakobsen:2021zvh}.

The full 5PM integrand is generated using the Berends-Giele type recursion relation discussed in ref.~\cite{Jakobsen:2023ndj},
and sorted into five self-force (SF) sectors according to their scaling with the masses $m_{1}$ and $m_{2}$:
\begin{align}\label{eq:sfExpansion}
\Delta p^{(5) \mu}_1=  m_{1} m_{2} & \Bigl ( m_2^{4} \Delta p^{(5) \mu}_{\text{0SF}}
+  m_1 m_2^3 \Delta p^{(5) \mu}_{\text{1SF}}   \\
+m_1^2 m_2^2 &\Delta p^{(5) \mu}_{\text{2SF}} 
+ m_1^3 m_2 \Delta p^{(5) \mu}_{\widebar{\text{1SF}}} 
 + m_1^4 \Delta p^{(5) \mu}_{\widebar{\text{0SF}}}\Bigr) \nn \,.
\end{align}
The powers of the masses follow from the number of times a wordline is touched in a given graph.
Here we compute the sub-leading self-force (1SF) contributions $\Delta p^{(5) \mu}_{\text{1SF}}$
and $\Delta p^{(5) \mu}_{\widebar{\text{1SF}}}$, as well as 
reproducing the 0SF contributions $\Delta p^{(5) \mu}_{\text{0SF}}$ and $\Delta p^{(5) \mu}_{\widebar{\text{0SF}}}$ that
follow from the geodesic motion in a Schwarzschild background.
The resulting integrand is reduced to scalar integrals via tensor reduction,
and ``planarized'' using partial fraction (eikonal) identities as described in ref.~\cite{Driesse:2024xad}.
In summary, all integrals are mapped to the 5PM-1SF planar family ($\int_{\ell}\coloneqq\int d^{D}\ell/(2\pi)^{D}$,
$\dd(x)\coloneqq2\pi\delta(x)$)
\begin{subequations}\label{eq:planarFamily}
\begin{align}
  {\cal I}_{\{n\}}^{\{\sigma\}}
  =
  \int_{\ell_1\cdots\ell_4}
  \frac{
    \dd^{(\bar{n}_1\!-\!1)}(\ell_1\cdot v_1)\prod_{i=2}^L\dd^{(\bar{n}_i\!-\!1)}(\ell_i\cdot v_2)
    }{
    \prod_{i=1}^{4}D_{i}^{n_{i}}(\sigma_{i})
    \prod_{I<J}D_{IJ}^{n_{IJ}}
    }\,,
\end{align}
where $\{\sigma\}$ and $\{n\}$ denote causal $\i 0^+$ prescriptions
and integer powers of propagators, respectively.
The four worldline propagators $D_i(\sigma_i)$ appearing  
are ($i=2,3,4$)
\begin{align}
  D_1&=\ell_1\cdot v_2+\sigma_1\i 0^+\,, &
  D_{i}&=\ell_i\cdot v_1+\sigma_i \i 0^+\,,
\end{align}
and the 14 gravitons propagators $D_{IJ}$ with $I=(0,1,i,q)$ are
\begin{align}\label{DIJs}
D_{1j}&=(\ell_1-\ell_j)^2 + \sigma_{4+j}\text{sign}(\ell_1^{0}-\ell_j^{0}) \i0^{+},   D_{q1}=(\ell_1+q)^2\nn\\
  D_{ij}&=(\ell_i-\ell_j)^2
  \,,
  D_{qi}=(\ell_i+q)^2
  \,,
  D_{01}=\ell_1^2
  \,,
  D_{0i}=\ell_i^2
  \,.
\end{align}
\end{subequations}
There are at most three bulk graviton propagators $D_{1i}$ that may go on-shell at 5PM order. 

\sec{IBP reduction}
Integration-by-parts (IBP) identities~\cite{Tkachov:1981wb,Chetyrkin:1981qh,Laporta:2000dsw} are used to 
reduce to master integrals. We employ a future release of {\tt KIRA 3.0} \cite{Maierhofer:2017gsa,Klappert:2020nbg} adapted to our needs that utilizes the {\tt FireFly} \cite{Klappert:2019emp,Klappert:2020aqs} library for reconstructing rational functions through finite-field sampling. 
We have 45 top-level sectors in the 5PM-1SF family that have been described in ref.~\cite{Driesse:2024xad}.
The integrals encountered in the planar family
\eqref{eq:planarFamily} have propagator powers in the range $n_{i/IJ}\in [-9,8]$. The strategies applied to reduce the run-time of
the IBP reduction are comparable to the conservative case \cite{Driesse:2024xad}. The final set of needed IBP replacement rules to master integrals generated comprises about 30GByte of data, and 
can be made available upon request.

\sec{Differential equations and function space}
The method of differential equations~\cite{Kotikov:1990kg,Gehrmann:1999as,Henn:2013pwa} is employed, where the matrices in 
 \eqref{eq:diffEqn} depend on the parameters $x$ and the dimensional regulator $\epsilon=(4-D)/2$. We take the physical limit $\epsilon \rightarrow 0$ to compute our observables. Therefore, we need the solutions of the integrals expanded in $\epsilon$. To systematically compute this expansion, we transform \eqref{eq:diffEqn} into canonical form~\cite{Henn:2013pwa} such that the $\epsilon$ dependence is factored out of the differential equation matrix:
\begin{align}\label{eq:canEqn}
	\frac{\d}{\d x} \vec{J}(x,\epsilon) = \epsilon \hat{A}(x)\, \vec{J}(x,\epsilon)\,,
	\end{align}
with $\vec{J}(x,\epsilon)=\hat T(x,\epsilon) \vec{I}(x,\epsilon)$ and $\epsilon \hat{A}(x) = (\hat{T}(x,\epsilon)\hat{M}(x,\epsilon)+\mathrm d\hat{T}(x,\epsilon)) T(x,\epsilon)^{-1}$.
The solution is then a path-ordered matrix exponential:
\begin{align}
	\label{eq:solutionDE}
	\vec{J}=\mathcal{P} e^{\epsilon \int_1^x dx \hat{A}(x)} \vec{j} \,,
\end{align}
where $\vec{j}$ encodes the boundary values of our integrals at $x=1$.

We take a bottom-up approach to determine the required transformation $\hat T(x,\epsilon)$.
For this, we sort our integrals into groups sharing the same set of propagators.
These so-called sectors are ordered from lower (fewer propagators) to higher (more propagators),
resulting in the block-diagonal matrix in Fig.~\ref{fig-4}.
We begin by $\epsilon$-factorizing the lower sectors and then move on to the higher sectors.
First, we transform the diagonal blocks, which are identified with the maximal cuts~\cite{Primo:2016ebd},
into $\epsilon$-form and then proceed to the off-diagonal contributions\footnote{IBP reductions were fully completed for all integrals before the diagonal blocks were identified.}.
Particularly for handling sectors coupled to the CY3 diagonal sector,
it is important to choose a good initial basis of integrals such that the relevant couplings are as simple as possible.
As we proceed to canonicalize, we adapt and improve our choice of initial basis accordingly.

The simplest diagonal blocks to canonicalize are those containing only multiple polylogarithms~\cite{Remiddi:1999ew, Goncharov:2010jf, Duhr:2011zq}, depicted in lilac in Fig.~\ref{fig-4}.
The algorithm \texttt{CANONICA} \cite{Meyer:2017joq} finds the necessary transformation to $\epsilon$-form for these blocks by making a suitable ansatz.
It is noteworthy that the complexity of this transformation, as well as the runtime, highly depends on the choice of initial integrals.
In general, we pick our initial basis integrals so that the regulator $\epsilon$
does not appear non-trivially in the denominators of the differential equation \eqref{eq:diffEqn}.

Diagonal blocks containing a K3 surface, depicted in purple in Fig.~\ref{fig-4}, are handled using \texttt{INITIAL} \cite{Dlapa:2020cwj}.
The \texttt{INITIAL} algorithm requires a pure\footnote{In this context, a pure integral is given as a linear combination of iterated integrals~\eqref{eq:integralDef} having no non-trivial pre-factors in $x$. For non-pure integrals, these pre-factors are non-trivial functions and are known as leading singularities.}
seed integral to construct an $\epsilon$-factorized differential equation through an ansatz tailored to the specific seed integral.
For each K3 sector we find an appropriate seed integral by analyzing the diagonal block at $\epsilon=0$.
We decouple each diagonal K3 block by switching to a derivative basis:
$\vec{I}_\text{K3}=(I_1,I_2,I_3)\rightarrow (I_1,I_1',I_1'')$ or $\vec{I}_\text{K3}=(I_1,I_2,I_3,I_4)\rightarrow (I_1,I_1',I_1'',I_4)$
depending on the size of the block, where $I_i$ are the master integrals of this sector ---
$I_4$ is chosen so that it decouples from $I_1$ and its derivatives as much as possible.
The choice of $I_1$ ensures that its third-order differential (Picard-Fuchs) equation ($\hat\theta=x\frac{\d}{\d x}$), 
\begin{align}
	\label{eq:PFK3}
	\left[  \hat\theta^3 -2 x^2 \left(2+4 \hat\theta +3 \hat\theta ^2+\hat\theta ^3\right)+x^4 (2+\hat\theta )^3   \right]\!\!\left. I_1\right|_{\eps=0} = 0\,,
\end{align}
has the explicit solution $\left. I_1\right|_{\eps=0} \propto \varpi_\text{K3}=(\frac{2}{\pi})^2 K^2(1-x^2)$, i.e.~it is proportional to a K3 period.
Unlike the polylogarithmic case, this third-order differential equation is not factorizable into first-order equations.
Using the normalized integral $I_1/\varpi_\text{K3}$ as seed,
\texttt{INITIAL} may then construct an $\epsilon$-form for the diagonal parts of our K3 sectors (similar to the 4PM case).

To canonicalize the single diagonal CY3 block, depicted in red in Fig.~\ref{fig-4},
we follow the discussion in ref.~\cite{Klemm:2024wtd}.
Similar to K3,
we first pick a suitable starting integral and make a basis change $\vec{I}_\text{CY3}=(I_1,I_2,I_3,I_4)\rightarrow (I_1,I_1',I_1'',I_1''')$.
The starting integral $I_1$ is chosen so that, when $\epsilon=0$,
it satisfies the Picard-Fuchs equation~\eqref{CY3},\footnote{Notice that the solutions of the Picard-Fuchs equation~\eqref{CY3} are given in terms of hypergeometric functions, e.g. $x\, {}_4\!F_3[\frac12,\frac12,\frac12,\frac12; 1, 1, 1 ; x^4]$. This implies that the integral $I_1$ for $\epsilon=0$ is a linear combination of hypergeometric functions.}
 thus determining the periods of our CY3 geometry.
From this equation and its four fundamental solutions $\varpi_0,...,\varpi_3$,
which we collect in the Wronskian matrix $W(x)=(\partial ^j \varpi_i)$ with $0\leq i,j,\leq 3$, we construct in multiple steps the rotation matrix into $\epsilon$-form.
This approach was invented in ref.~\cite{Gorges:2023zgv} and further developed in ref.~\cite{Neganew},
and in the K3 case, it is equivalent to the \texttt{INITIAL} algorithm.
The process involves the following three steps:
\begin{enumerate}
	\item We split the Wronskian matrix into a semi-simple and unipotent part $W=W^\text{ss}\cdot W^\text{u}$. Naively, one can understand this splitting as a decomposition of the maximal cuts of the CY3 sector into their leading singularities and pure parts. The unipotent part is named after the unipotent differential equation it fulfills:
\begin{align}
	(\mathrm d-A^\text{u}(x))W^\text{u}(x)=0 \,,
\end{align}
where $A^\text{u}(x)$ is nilpotent. The matrix $A^\text{u}(x)$ can generally be written in terms of invariants, known as $Y$-invariants, of the Calabi-Yau variety and was explicitly given for our CY3 in ref.~\cite{Klemm:2024wtd}.
\item We rotate our integrals with the inverse of $W^\text{ss}$, which strips them of their leading singularities.
(The analagous operation for K3 is 
normalizing $I_1$ with the K3 period $\varpi_\text{K3}$, leaving only the pure part.)
To parametrize all degrees of freedom in $W^\text{ss}$ of a CY3, one needs the holomorphic solution $\varpi_0$, an additional function
\begin{align}
	\alpha_1	=	\frac{\varpi_0^2}{x(\varpi_0\varpi_1'-\varpi_0'\varpi_1)}\,,
\end{align}
called the structure series, and their derivatives.
The appearance of $\alpha_1$ is new compared to the K3 case,
and shows the increased complexity in structure of a CY3.
In ref.~\cite{Bogner:2013kvr}, the structure series $\alpha_1$
(and more generally the $Y$-invariants) are generally defined,
and used to construct a normal form of a CY differential equation;
more specifically for our CY3 they were derived in ref.~\cite{Klemm:2024wtd}.
For $\epsilon$-factorizing our differential equations,
it is important that this normal form of the CY differential equation is in a factorized form with respect to its derivatives.
To eliminate all redundancies in this step, we must utilize Griffiths transversality ---
an essential property of CY geometries that yields quadratic relations between their periods --- to simplify the form of $W^\text{ss}$.
\item After completing these steps and appropriately rescaling in $\epsilon$ to arrange the weights of the integrals, the diagonal block of our CY3 looks like:
\begin{align}
\begin{aligned}
	\frac{\mathrm d}{\mathrm dx} \vec{\tilde I}_\text{CY3}	&=\sum_{i=-2}^1\epsilon^i \tilde M^i_\text{CY3}(x)\vec{\tilde I}_\text{CY3}\,, \quad\text{with} \\
	\vec{\tilde I}_\text{CY3}	&=	T_{\epsilon\text{-scalings}} \left(W^\text{ss}\right)^{-1} \vec{I}_\text{CY3}	\, .
\end{aligned}
\end{align}
We find an $\eps$-form by acting with a suitable set of transformation matrices on $\vec{\tilde I}_\text{CY3}$,
working order-by-order in $\eps$ starting from  $\epsilon^{-2}$.
The process requires us to introduce four new functions $G_i(x)$ ($i=1,\ldots, 4$),
which obey a first-order differential equation containing $\varpi_0, \alpha_1$, their derivatives, and $G_j(x)$ functions with $j<i$.
For example,
\begin{align}
	\qquad \ G_1'(x) =-\frac{96 x \left(x^4+1\right) \varpi_0(x)^2}{(x-1)^2 (x+1)^2 \left(x^2+1\right)^2 \alpha_1(x)} 	\, .
\end{align}
Because of this structure, the functions $G_i(x)$ are all expressible as iterated integrals of Calabi-Yau periods and associated functions. These functions were previously introduced in terms of a different variable in ref.~\cite{Klemm:2024wtd}, and are listed for our conventions in the Supplementary Information of this article.
\end{enumerate}
We now have the $\epsilon$-form of the diagonal part of the CY3 sector,
and thus a canonical form of all diagonal blocks.
We refer to this intermediate basis, wherein all diagonal blocks are in $\eps$-form, as $\mathfrak J$.
The next stage in canonicalization involves tackling the off-diagonal blocks.
To do so, we distinguish between off-diagonal blocks coupled to the CY3 sector, which require special care, and the rest.

We have developed our own algorithm to transform the off-diagonal entries of our differential equation
that do not couple to the CY3 sector but can depend on K3 functions.
This algorithm utilizes \texttt{FINITEFLOW}~\cite{Peraro:2019svx} and \texttt{MultivariateApart}~\cite{Heller:2021qkz},
and provides suitable ans\"atze also including elliptic contributions for the required transformations.
It is similar to algorithms used for polylogarithmic off-diagonals, such as those found in \texttt{CANONICA} or \texttt{Libra}~\cite{Lee:2020zfb}.

For sectors polylogarithmic or K3 on their diagonal blocks, yet also coupled to the CY3 sector,
a good intial basis of integrals is essential in order to minimize these couplings.
One possibility to identify such candidates is to perform an integrand analysis,
usually done in the Baikov representation~\cite{Frellesvig:2023bbf,Frellesvig:2024zph} of the integrals.
However, we instead found it simpler to use the diagonals themselves in order to derive constraints on the initial choice of integrals,
expanding upon the ideas of ref.~\cite{Dlapa:2023hsl}.
Having now found canonical masters on the diagonals, i.e.~the maximal cuts,
our strategy is to choose initial candidate integrals that are as closely as possible related to these canonical masters within their respective diagonal blocks.
More precisely, we search for candidates that, on their diagonal blocks, are given by a linear combination of the canonical maximal cut integrals and overall functions of $\epsilon$ and $x$:
\begin{align}
	I_i^\text{candidate}  = f(\epsilon)g(x) \sum_k c_k\,  \mathfrak J_k \, ,
\label{eq:maxcutrel}
\end{align}
where the $c_k$ are constant numbers.

We expect that such a ``good'' choice of candidate integrals only requires minimal corrections to form a canonical basis.
For certain sectors that are polylogarithmic on their corresponding diagonal block,
we need to enlarge these types of constraints by combining different polylogarithmic sectors and requiring Eq.~\eqref{eq:maxcutrel} to hold beyond a single diagonal block.
Thus, we obtain further conditions resulting from off-diagonal couplings between separate polylogarithmic blocks.
In some instances, we can also relax condition \eqref{eq:maxcutrel} by considering the $c_k(x)$ as functions of $x$ and still find easy transformations.
The use of IBPs makes this procedure efficient,
and allows us to find all transformations for the coupling to the CY3 manually, proceeding similarly as for the diagonal of the CY3 sector.
We build successive transformations, removing iteratively all non-linear-in-$\epsilon$ contributions starting with the highest negative power of $\epsilon$.
By doing so, for some integrals we need to introduce sixteen new functions $G_5(x),\hdots, G_{20}(x)$,
which again satisfy first-order differential equations listed in the supplementary information.
More specifically, for the mixings between K3 and CY3 sectors,
we introduce new functions whose first-order differential equations contain the periods of both the K3 and the CY3. For example,
\begin{align}	
	G_8'(x)	=	\frac{\varpi_\text{K3}(x) G_3(x) \varpi_{0}(x) \alpha_1 '(x)}{\alpha_1(x)^2}	\, .
\end{align}
This concludes the canonicalization process of the whole differential equation system.

Having converted our matrix into its canonical form,
$\hat A(x)$ provides all integration kernels needed to express the master integrals as iterated integrals~\eqref{eq:integralDef}.
We selected a set of linearly independent kernels by examining their small velocity expansion.
Our observables consist of iterated integrals that include K3, CY3, and mixed integration kernels, functions from the rotation matrix $\hat T(x,\epsilon)$,
and algebraic functions from the decomposition in terms of initial master integrals.
We need at most four-times-iterated integrals; all multiple polylogarithms are constructed from the kernels
$\{\frac 1 x , \frac 1 {1\pm x},\frac{x}{ 1+x^{2}}\}$, and have a maximum transcendental weight of 3.

\sec{Boundary fixing}
A complete solution to the differential equation \eqref{eq:diffEqn} requires the determination of integration constants in the form of
boundary integrals, i.e.~master integrals in the static limit $x\to 1$ ($v\to 0$) which are functions only of $\epsilon$.
As the integration and $x\to1$ limit do not commute,
we use the method of regions~\cite{Beneke:1997zp,Smirnov:2012gma,Becher:2014oda} to isolate contributions with definite ($\eps$-dependent)
scalings in the velocity $v$, and series expand integrals at the level of the integrand.
These so-called regions are associated with different velocity scalings of the bulk graviton momenta $\ell_i$,
which can be either potential (P) or radiative (R):
\be\label{PRregions}
	  \ell^\text{P}_i=(\ell_i^0,\Bell_i)\sim(v,1)\,,  \qquad
	  \ell_i^\text{R}=(\ell_i^0,\Bell_i)\sim(v,v)\, .
\ee
There are three propagators $\{ D_{12},D_{13},D_{14}\}$ that may enter both regions;
the rest are kinematically restricted to P (including the velocity-suppressed P: $(v^2,v)$) by the presence of energy-conserving delta functions $\delta(\ell_i\cdot v_j)$.
We thus denote the four possible regions as (PPP), (PPR), (PRR) and (RRR).
We needed to evaluate 14+14 (even+odd) boundary integrals in the (PPR), 5+5 in the (PRR) and 4+4 in the (RRR) sector;
in addition to the 28+18 (PPP) boundary integrals that were already determined in the conservative case~\cite{Driesse:2024xad}
(here there is no distinction between Feynman and retarded bulk propagators).
We perform all integrals analytically and check them numerically using \texttt{pySecDec} \cite{Borowka:2017idc,Borowka:2018goh,Heinrich:2021dbf,Heinrich:2023til,Vermaseren:2000nd,Ruijl:2017dtg} and \texttt{AMPred} \cite{Chen:2019mqc,Chen:2019fzm,Chen:2020wsh,Chen:2024xwt}.

All new boundary integrals for us to evaluate, as compared with ref.~\cite{Driesse:2024xad},
contain radiative graviton propagators.
Our main strategy is to perform them via Schwinger parametrization,
wherein we also benefit by explicitly integrating out loops involving only gravitons, leaving lower-loop integrals. In doing so, we evaluate at most two-loop integrals in all but two cases. 
These two are genuine three-loop integrals for which we did not manage the integration over Schwinger parameters, and therefore moved over to the time domain in order to establish identities at the level of the series coefficients. 
In general, our integrations yield generalized hypergeometric functions ${}_pF_q$,
which can be series expanded in $\eps$ by numerically expanding them to high precision and
reconstructing analytic expressions using an ansatz and an integer relation algorithm.
Expressions for all boundary integrals, and our methodology for deriving them,
are elaborated in the Supplementary Information.

\sec{Results}
Our main results are full expressions --- including dissipation ---
for the 1SF and $\widebar{\rm 1SF}$ parts of the momentum impulse,
$\Delta p_{\rm 1SF}^{(5)\mu}$ and $\Delta p_{\rm \widebar{1SF}}^{(5)\mu}$~\eqref{eq:sfExpansion}.
They are expanded on basis vectors:
\begin{subequations}\label{eq:impulse}
\begin{align}
	\Delta p_{\rm{1SF}}^{(5)\mu}
	&=
	\frac1{b^5}
	\Big(
		\hat b^\mu c_b(\gamma)
		+
		\check v_2^\mu 
		c_v(\gamma)
		+
		\check
		v_1^\mu
		c_v'(\gamma)
		\Big)
		\,,\\
  \Delta p_{\rm\widebar{1SF}}^{(5)\mu}
	&=
	\frac1{b^5}
	\Big(
		\hat b^\mu \bar c_b(\gamma) 
		+
		\check v_2^\mu 
		\bar c_v(\gamma) 
		+
		\check
		v_1^\mu
		\bar c_v'(\gamma)
		\Big)
		\,,
\end{align}
\end{subequations}
also pulling out an overall factor of the impact parameter $b$.
This decomposition constitutes a split into parts originating from integrals of even and odd parity in $v^\mu_i$.
The basis vectors are the impact parameter $\hat b^\mu = (b_2^\mu-b_1^\mu)/b$ and dual velocity vectors
$\check v_1^\mu =(\gamma v_2^\mu-v_1^\mu)/(\gamma^2-1)$ and $\check v_2^\mu =(\gamma v_1^\mu-v_2^\mu)/(\gamma^2-1)$.
The coefficients of main interest here are $\{c_b,\bar{c}_b\}$ and $\{c_v,\bar{c}_v\}$, because the set
$\{c'_v,\bar{c}'_v\}$ being determined by lower-PM results using preservation of mass $p_1^2=(p_1+\Delta p_1)^2$.
The coefficients are further decomposed into those
with an even or odd number of radiative gravitons in the boundary fixing:
\begin{subequations}
\begin{align}
	c_w(\gamma) =
	c_{w,{\rm even}}(\gamma) 
	+
	c_{w,{\rm odd}}(\gamma) 
	\,,\\
  \bar{c}_w(\gamma) =
	\bar{c}_{w,{\rm even}}(\gamma) 
	+
	\bar{c}_{w,{\rm odd}}(\gamma) 
	\,,
\end{align}
\end{subequations}
with $w\in\{b,v\}$.
The even part is defined from $\rm(PPP)+(PRR)$, and the odd part from $\rm(PPR)+(RRR)$.
The two parts have distinctive parities under flipping the sign of the relative velocity $v\to-v$.
Even and odd sectors then give rise to integer and half-integer post-Newtonian (PN) orders, respectively.

The 1SF and $\rm \widebar{1SF}$ parts of the impulse therefore each consist of 
four non-trivial coefficients, labeled by $b$ or $v$,
each having an even or odd number of radiative gravitons $\rm R$.
We expand each of these in sets of basis functions $F(\gam)$ with coefficient functions
 $d(\gam)$, being polynomial
up to factors of $(\gam^2-1)$
\begin{align}
\begin{aligned}
	c_{w,z}(\gam)
	&=
	\sum_\alpha 
	d^{(\alpha)}_{w,z}(\gam)
	F^{(\alpha)}_{w,z}(\gam) 
	\\
	&+
	\sum_\alpha
	d_{w,z}^{(\alpha,\rm tail)}(\gamma)
	F_{w,z}^{(\alpha,\rm tail)}(\gamma)
	\log(\gamma-1)
	\,,
\end{aligned}
\end{align}
barred coefficients being expanded in the same way.
Here, $z\in\{\text{even},\text{odd}\}$ counts the parity of radiative gravitons.
The second line with a logarithm produced from the cancellation of $1/\eps$ poles between the different boundary regions is associated with tails. It is relevant for all coefficients except $c_{b,\rm odd}$, $\bar c_{b,\rm odd}$ and $c_{v,\rm even}$.
The b-type basis functions are all multiple polylogarithms of maximum weight three and, thus, relatively simple.
In contrast, the v-type basis functions are much more complex.
They generally have the structure of Eq.~\eqref{eq:integralDef} with kernels depending on the Calabi-Yau periods.
All basis functions and polynomial coefficients are provided in the ancillary file in our repository submission~\cite{zenodo}.

The total loss of four-momentum at 5PM order, using Eq.~\eqref{Prad},
is given schematically by (see e.g. ref.~\cite{Bini:2022enm})
\begin{align}\label{Pradfinal}
	P_{\rm rad}^{(5)\mu}
	&=
	\frac{ M^6\nu^2}{b^5}
	\Big(
		\big[
			r_1(\gam) \hat b^\mu
	+r_2(\gam)(v_1^\mu-v_2^\mu)
	\big]
		\frac{m_1-m_2}{M}
		\nonumber
		\\
		&+
		\big[ 
			v_1^\mu+v_2^\mu
			\big] (
			r_3(\gam) 
			+\nu\, r_4(\gam)
		)
	\Big)
	\,.
\end{align}
Our 1SF/$\rm \widebar{1SF}$ result fixes all coefficients except $r_4(\gam)$.
Similarly, we may derive the relative scattering angle\cite{Bini:2021gat,Jakobsen:2023pvx} $\theta=\arccos(\mathbf{p}_{\rm in}\cdot\mathbf{p}_{\rm out}/|\mathbf{p}_{\rm in}||\mathbf{p}_{\rm out}|)$.
Here $\mathbf{p}_{\rm in}=\mathbf{p}_1=-\mathbf{p}_2$ is the incoming momentum in the centre-of-mass frame, and
\begin{equation} 
	\mathbf{p}_{\rm out}=\mathbf{p}_{\rm in}+\Delta \mathbf{p}_1 + \frac{E_1}{E}\mathbf{P}_{\rm recoil} + \cO(G^6)
	\,,
\end{equation}
is the (relative) outgoing momentum.
We expand the scattering angle in $G$: 
\begin{align}
\theta
&=
\Gamma
\sum_{n=1}^5
\bigg(\frac{GM}{b}\bigg)^n
\sum_{m=0}^{\lfloor\frac{n-1}{2}\rfloor}
\nu^m \theta^{(n,m)}(\gamma)
\nonumber
\\
&\ \ \ \ \ +
\frac1{\Gam}
\sum_{n=4}^5
\bigg(\frac{GM}{b}\bigg)^n
\nu^{n-2} \theta^{(n,n-2)}(\gamma)
+\cO(G^6) \, .
\end{align}
Both the angle and $P_{\rm rad}^{(5)\mu}$ expansion coefficients are 
separated into parts even and odd under $v\to-v$, and expanded on suitable basis functions.
All of these results can be found in the {\tt observables.m}
ancillary file in our repository submission\cite{zenodo}.

\sec{Checks}
Our result for the impulse has been checked internally by the cancellation of $1/\eps$
poles and obeying the mass condition $p_1^2=(p_1+\Delta p_1)^2$ ---
verifying the $\{c'_v,\bar{c}'_v\}$ coefficients~\eqref{eq:impulse}.
We have also checked that the (PPP) 1SF contribution is related to its $\rm \widebar{1SF}$ counterpart by symmetry.
In addition, we have performed several checks using the non-relativistic ($v\to 0$) limit.
First, PN results for the relative scattering angle at 5PM and to first order in self-force are known to 5.5PN order~\cite{Bini:2021gat,Bini:2022enm,Bini:2020hmy}.
These include conservative terms at integer PN orders, starting from 0PN order which were already matched in ref.~\cite{Driesse:2024xad},
plus dissipative terms appearing at 2.5PN, 3.5PN, 4PN, 4.5PN, 5PN and 5.5PN orders which we reproduce.
Furthermore, and in the same works~\cite{Bini:2021gat,Bini:2022enm,Bini:2020hmy},
dissipative PN results for the radiation of energy and recoil of the two-body system were reported at a relative 3PN order to their leading order.
These allow us to perform non-trivial checks on $r_1(\gamma)$, $r_2(\gamma)$ and $r_3(\gamma)$ of \eqref{Pradfinal} to relative 3PN order.

As an example of our results,
we print explicitly the series expansion in $v$ of the $G^5$ component of $E_{\rm rad}$:
\begin{align}
	E_{\rm rad}^{(5)}
	&=\frac{
		 M^6
		 \nu^2
		 }{\Gamma b^5}
	\Big[
		(1-\gamma )r_2(\gamma)+(1+\gamma)r_3(\gamma)
		+\cO(\nu)
		\Big]\nonumber \\
	&=
	\frac{M^6\nu^2\pi}{5\Gamma b^5 v^3}
	\Big[
		122
		+
		\frac{3583}{56}
		v^2
		+
		\frac{297\pi^2}{4}
		v^3
		-
		\frac{71471}{504}
		v^4
		\nonumber	\\&+
		\Big(\frac{9216}{7}-\frac{24993\pi^2}{224}\Big)v^5
		\nonumber
		\\&
		+
		\Big(
		\frac{2904562807}{6899200}
		+\frac{99\pi^2}{2}
		-
		\frac{10593}{70} \log{\frac{v}{2}}
		\Big)v^6
		\nonumber
		\\&
		+\Big( 
			\frac{7296}{7}
			-
			\frac{2927\pi^2}{28}
			\Big)v^7
			\nonumber
		\\&
		+\Big( 
			\frac{4924457539}{29429400}
			+
			\frac{8301\pi^2}{112}
			-
			\frac{491013}{3920}
			\log{\frac{v}{2}}
			\Big)v^8
			\nonumber
			\\&
			+\Big( 
				\frac{99524416}{40425}
				\!-\!
				\frac{46290891\pi^2}{157696}
				\Big)v^9
		\!+\!\cO(v^{10}\!,\nu)
		\Big]
		\ .
\end{align}
The terms in the square brackets up to $v^6$ reproduce the known PN results of $E_{\rm rad}$
with the remaining three lines providing hitherto unknown further terms.
Naturally, this series expansion can be extended to any order in $v$ using our present results.
Explicit results relevant for the PN checks of the relative scattering angle and recoil are given in the Supplementary Information.

\section*{Data availability}

The complete set of observables up to
the 5PM-1SF order 
 of our work, the scattering angle, radiated energy and impulse four-vector,
 are available in the {\tt zenodo.org} open repository submission 
\href{https://doi.org/10.5281/zenodo.14604438}{https://doi.org/10.5281/zenodo.14604438} \cite{zenodo} accompanying this article. They are provided
in the form of {\tt Mathematica} computational notebooks.

\section*{Code availability}

The analytical results of our work employed the commercial computer algebra
system {\tt Wolfram Mathematica}\cite{Mathematica} as well as the freely available computer algebra system
 {\tt Form}~\cite{Ruijl:2017dtg}. 
 The publicly available programs for reconstructing rational functions through finite-field sampling, \texttt{FireFly}\cite{Klappert:2019emp,Klappert:2020aqs} and \texttt{FiniteFlow}\cite{Peraro:2019svx}, were utilized. Additionally, programs designed for the analytical treatment of DEs, including \texttt{MultivariateApart}\cite{Heller:2021qkz}, \texttt{CANONICA}\cite{Meyer:2017joq}, and \texttt{INITIAL}~\cite{Dlapa:2020cwj}, were employed as described in the Methods section.
Crucially, the IBP software package {\tt KIRA 3.0}\cite{Maierhofer:2017gsa,Klappert:2020nbg}  was used for the reduction to master integrals in its beta
version that can be made accessible by JU upon reasonable request.
For the numerical checks of the boundary integrals the public software 
\texttt{pySecDec} \cite{Borowka:2017idc,Borowka:2018goh,Heinrich:2021dbf,Heinrich:2023til,Vermaseren:2000nd,Ruijl:2017dtg} and \texttt{AMPred} \cite{Chen:2019mqc,Chen:2019fzm,Chen:2020wsh,Chen:2024xwt} was
employed.
All our code and observables are provided in {\tt Mathematica} computational
notebooks: The WQFT Feynman rules necessary to perform the 5PM computation, the
explicit form of the boundary integrals as well as the 
 20 transcendental $G_{i}$ functions
 necessary to derive the canonical form of the differential equations. These computational notebooks are freely available in the {\tt zenodo.org} repository submission 
\href{https://doi.org/10.5281/zenodo.14604438}{https://doi.org/10.5281/zenodo.14604438} \cite{zenodo}. 
The set of IBP replacement rules to the  master integrals (30 GB of code) 
generated  upon using {\tt KIRA 3.0} can be made available 
upon reasonable request to the corresponding author.

\section*{Author Contributions}

 As is customary in high-energy physics, the authors and their contributions are listed in alphabetical order.
MD, GUJ, GM, JP and BS performed the integrand generation and planarization;
MD, BS and JU conducted the IBP reduction, CN and BS canonicalized the differential equations;
AK, CN and BS identified the underlying Calabi-Yau structures;
MD, GUJ, GM, JP, BS and JU contributed to the boundary integrals and boundary matching;
GUJ, GM and BS performed the checks with the existing post-Newtonian results;
MD, GUJ, CN and BS simplified the final results and implemented them numerically;
GM and JP secured the funding and provided leadership of the project.
All authors contributed to the writing of the manuscript.

\section*{Acknowledgements}
We thank Martin Beneke, Wen Chen, Christoph Dlapa, Roger Morales, Rafael Porto, Nikolaos Syrrakos, Lorenzo Tancredi, Duco van Straten, and Matthias Wilhelm for insightful discussions and J.~Kleinmond for providing the waveform visualization of Fig.~1. 
This work was funded by the Deutsche Forschungsgemeinschaft
(DFG, German Research Foundation)
Projektnummer 417533893/GRK2575 ``Rethinking Quantum Field Theory'' (GUJ, GM, BS, JP, JU) 
and 508889767/FOR5372/1 ``Modern Foundations of Scattering Amplitudes'' (AK, JP),
and in part by the Excellence Cluster ORIGINS (CN)
under Germany’s 
Excellence Strategy – EXC-2094-390783311. This work was funded
by the UK Royal Society under grant URF\textbackslash R1\textbackslash 231578,
``Gravitational Waves from Worldline Quantum Field Theory'' (GM).
This work was funded by the European Union through the 
European Research Council under grant ERC Advanced Grant 101097219 (GraWFTy) (MD, JP) and ERC Starting Grant 949279 (HighPHun) (CN).
Views and opinions expressed are, however, those of the authors only and do not necessarily reflect those of the European Union or European Research Council Executive Agency. Neither the European Union nor the granting authority can be held responsible for them.
This research was supported by the Munich Institute for Astro-, Particle and BioPhysics (MIAPbP), which is funded by the Deutsche Forschungsgemeinschaft (DFG, German Research Foundation) under Germany's Excellence Strategy – EXC-2094 – 390783311 (GUJ, GM, JP, BS).
The authors gratefully acknowledge the computing time made available to them on the high-
performance computer Lise at the NHR Center Zuse-Institut Berlin (ZIB). This center is jointly supported by the Federal Ministry of Education and Research and the state governments participating in the National High-Performance Computing (NHR) joint funding program 
(http://www.nhr-verein.de/en/our-partners).

\section*{Competing interests}
The authors declare no competing interests.

\begin{widetext}
\newpage
\setcounter{figure}{2} 

\renewcommand{\theequation}{S.\arabic{equation}}

\centerline{\bf SUPPLEMENTARY INFORMATION}


\section{Further Details of Results}
The main result of our work is the 5PM-1SF component of the momentum impulse $\Delta p_1^\mu$,
from which the corresponding components of the scattering angle $\theta$, loss of energy $E_{\rm rad}$ and recoil $\mathbf{P}_{\rm recoil}$ may subsequently be derived.
The 5PM-1SF impulse  is parametrized by eight coefficients:
$c_{b,\rm even/odd}(\gam)$, $c_{v,\rm even/odd}(\gam)$, $\bar c_{b,\rm even/odd}(\gam)$ and $\bar c_{v,\rm even/odd}(\gam)$,
plus the simpler $c_v'(\gam)$ and $\bar c_v'(\gam)$ coefficients that are pre-determined by lower-PM results.
In Eq.~(\ref{23}), the former coefficients were expanded in terms of
$\gamma$-dependent functions $F_{b/v,\rm even/odd}(\gam)$ multiplied by polynomials $d_{b/v,\rm even/odd}(\gam)$.
The functions $F_{b,\rm even/odd}$ are simpler than $F_{v,\rm even/odd}$
due to the absence of K3 and CY3 periods in the former,
the only integration kernels within the iterated integrals~(4) being $x^{-1}$, $(1\pm x)^{-1}$ and
$x/(1+x^2)=((x+i)^{-1}+(x-i)^{-1})/2$ --- i.e.~simple poles.
These functions may therefore be expressed in terms of multiple polylogarithms (MPLs):
\begin{align}
  G(a_1,\dots,a_n;y)
  =
  \int_0^{y}
  \frac{
    \d y'
  }{y'-a_1}
  G(a_2,\dots,a_n;y')
  \, .
\end{align}
In order to apply a conventional definition of MPLs,
for convenience here we have introduced $y=1-x$.
The letters $a_i$ are then shifted to $\{0,1,2,1\pm\ii\}$,
with the two imaginary letters occurring always in pairs.
All expansion functions of the $b$ coefficients $F_{b,\rm even/odd}(\gam)$
(and their $\widebar{\rm 1SF}$ counterparts) are linear combinations of MPLs of this kind up to weight 3.

Besides these same MPLs (up to weight-3),
the functions $F_{v,\rm even}(\gam)$ (and $\widebar{\rm 1SF}$)
also involve the K3 period $\varK(\gam)$ and its derivatives $\varK'(\gam)$, $\varK''(\gam)$.
The K3 period also occurs within the following three iterated integrals~(4):
\begin{subequations}
\begin{align}
  &\mathcal{I}\left(\frac{1+x^2}{x}\varK(x);x\right)\,,\\
  &\mathcal{I}\left(\frac{1}{(1-x^2)x\,\varK(x)},\frac{1+x^2}{x}\varK(x);x\right)\,,\\  
  &\mathcal{I}\left(\frac{1}{(1-x^2)x\,\varK(x)},\frac{1}{(1-x^2)x\,\varK(x)},\frac{1+x^2}{x}\varK(x);x\right)\,.
\end{align}
\end{subequations}
Finally, $F_{v,\rm odd}(\gam)$ (and $\widebar{\rm 1SF}$) depend on both the K3 and CY3 periods,
$\varK(\gamma)$ and $\varCY(\gamma)$,
their derivatives $\varK(\gamma)$, $\varK'(\gamma)$, $\varCY'(\gamma)$, $\varCY''(\gamma)$ and $\varCY'''(\gamma)$,
plus $\alpha(x)$ and $\alpha'(x)$.
There are also now iterated integrals $\cal I$ whose kernels depend on the CY3 and K3 periods and $\alpha(x)$,
with up to four iterations.
Finally, we encounter all the $G_i(\gam)$ functions given below in Eqs.~\eqref{eq:GFunctions},
which are also iterated integrals on kernels that depend on the periods and $\alpha(x)$ and their derivatives.

In all coefficients except $c_{b,\rm odd}$, $\bar c_{b,\rm odd}$ and $c_{v,\rm even}$,
a logarithm $\log(\gam-1)$ is present in the results,
and no other functions have a logarithmic divergence in the limit $\gam\to1$.
When using MPLs, we choose basis functions that are free of logarithms in this limit.
For the iterated integrals of CY3 and K3 periods, however, we have no general procedure for identifying such logarithms.
Wherever possible,
we have collected iterated integrals together in linear combinations so that logarithms of $\gam-1$ cancel in the $\gam\to1$ limit,
but this was not possible in all cases.
When one collects all terms, however, such cancellations do occur.
Finally, and as a general rule,
functions $F_{b/v,\rm even/odd}(\gam)$ are chosen with a definite parity $v\to-v$ wherever possible.
This follows the number of radiative gravitons and is equivalent to $x\to1/x$ or $\sqrt{\gam^2-1}\to-\sqrt{\gam^2-1}$.
For the more advanced basis functions, with iterated integrals involving the CY3 and K3 periods,
we were not always able to construct basis functions with the symmetry property ---
a complete and systematic organization of these functions is left for future work.

\subsection{Scattering angle}
The 5PM component of the relative scattering angle depends only on the $b$-type coefficients of the 5PM impulse,
and may therefore be expressed solely in terms of MPLs ---
as was the case for its conservative even-in-$v$ part~\cite{Driesse:2024xad}.
The relevant coefficients of Eq.~(6),
determined by our 5PM $\rm 1SF$ and $\widebar{\rm 1SF}$ results,
are $\theta^{(5,1)}(\gam)$ and $\theta^{(5,3)}(\gam)$.
Their contribution from an even or odd number of radiative gravitons,
$\theta^{(5,1)}_{\rm even/odd}(\gam)$ and $\theta^{(5,3)}_{\rm even/odd}(\gam)$, are also, respectively, even and odd with respect to the symmetry $v\to-v$.
In a PN velocity expansion on has
\begin{subequations}
\begin{align}
  \theta^{(5,1)} &= 
  \frac{4}{5 v^8}-\frac{137}{5 v^6}+\frac{3008}{45 v^5}+\frac{\frac{41 \pi
   ^2}{4}-\frac{3427}{6}}{v^4}+\frac{\frac{84 \pi
   ^2}{5}-\frac{4096}{1575}}{v^3}+\frac{-\frac{12544 \log(2v)}{45}+\frac{3593 \pi
   ^2}{72}-\frac{445867}{432}}{v^2}+\frac{\frac{2144536}{11025}+\frac{453 \pi
   ^2}{35}}{v}
   \nonumber
   \\
   &+\left(-\frac{7552 \log(2v)}{1575}+\frac{246527 \pi
   ^2}{1440}-\frac{1111790903}{756000}\right)+\left(\frac{19424344}{363825}+\frac{1787
   \pi ^2}{672}\right) v
   \nonumber
   \\
   &+\left(-\frac{1762784 \log(2v)}{11025}-\frac{184881 \pi
   ^2}{2240}+\frac{56424801733}{49392000}\right)
   v^2+\left(\frac{16004496043}{104053950}-\frac{835619 \pi ^2}{59136}\right)
   v^3
   +\cO(v^4)\,,
  \\
  \theta^{(5,3)} 
  &=
  \frac{128}{3 v}+\left(\frac{14528}{175}+\frac{37 \pi ^2}{10}\right) v-\frac{22016
   v^2}{225}+\left(\frac{3262832}{33075}+\frac{893 \pi ^2}{112}\right) v^3+\frac{2877184
   v^4}{7875}+\left(\frac{15803 \pi ^2}{960}-\frac{4464536}{14553}\right)
   v^5
   \nonumber
   \\
   &+\frac{970766528 v^6}{1819125}+\left(\frac{11234077 \pi
   ^2}{394240}-\frac{258810752887}{780404625}\right) v^7+\frac{83694772064
   v^8}{70945875}
   \nonumber
   \\
   &+\left(\frac{119425757 \pi
   ^2}{2795520}-\frac{5667010769993}{5533778250}\right) v^9+\frac{11015320038116
   v^{10}}{5462832375}+
   \cO(v^{11})
  \,.
\end{align}
\end{subequations}
As discussed in Methods, the PN expansion of $\theta^{(5,1)}$ has been checked up to and including $v^0$ against known PN data. The complete expressions are provided in the
{\tt observables.m} file of the repository submission.
Here the MPLs may be PN expanded using conventional tools, such as { \tt PolyLogTools}~\cite{Duhr:2019tlz}, which is why we do not provide explicit $v$-expansions there.

\subsection{Radiated momentum}

The 5PM-1SF component of the loss of four-momentum $P_{\rm rad}^\mu$~(\ref{Pradfinal}) is given by the impulse coefficients via
\begin{subequations}
\begin{align}
  r_1(\gam)
  &=
  c_b(\gam)-\bar c_b(\gam) \, ,
  \\
  r_2(\gam) 
  &=
  \frac{c_v(\gam)-\bar c_v(\gam)-c'_v(\gam)+\bar c'_v(\gam)}{2(\gam-1)} \, ,
  \\
  r_3(\gam)
  &=-
  \frac{c_v(\gam)+\bar c_v(\gam)+c'_v(\gam)+\bar c'_v(\gam)}{2(\gam+1)} \, .
\end{align}
\end{subequations}
Thus, $r_1$ depends only on MPLs and $r_{2,3}$
have the full generality of $c_{v,\rm odd}(\gam)$ and $\bar c_{v,\rm odd}(\gam)$ .
These results are collected in the {\tt observables.m} file of the repository submission.
In the centre-of-mass frame, wherein $P_{\rm rad}^\mu=(E_{\rm rad},\mathbf{P}_{\rm recoil})$,
the 5PM components take the form ($\Gamma=\sqrt{1+2\nu (\gamma-1)}$, $\nu=\mu/M=m_{1}m_{2}/M^{2}$, $M=m_{1}+m_{2}$)
\begin{align}
  E_{\rm rad}^{(5)}
	&=
    \frac{M^6 \nu^2}{\Gamma b^5}
    \big[
     (1-\gamma) r_2(\gam)+(1+\gamma)r_3(\gam)
		+\cO(\nu)
    \big]\,,\\
  \mathbf{P}_{\rm recoil}^{(5)}
  &=
  \frac{M^5\nu^2(m_1-m_2)}{b^5}
  \Big(
    r_1(\gam) \mathbf{\hat b}
    +
    (r_2(\gam)-r_3(\gam))
    \frac{\mathbf{p}_{\rm in}}{\mu}
    +
    \cO(\nu)
    \Big)\,.
\end{align}
An explicit formula for $E_{\rm rad}^{(5)}$ up to $v^6$ was provided in Methods~(\ref{27}).
The PN-expanded components of the recoil are
\begin{align}
r_{1}(\gamma)&=
-\frac{64}{3 v^2}-\frac{16192}{525}-\frac{37 \pi ^2}{20}-\frac{30208 v}{225}+\left(-\frac{856768}{33075}-\frac{3429 
  \pi^2}{1120}\right) v^2-\frac{22016 v^3}{2625}+\left(-\frac{1117888}{4851}-\frac{80723 \pi ^2}{13440}\right) v^4
  \nn\\ &
  -\frac{123897344 v^5}{606375}+\left(-\frac{36746586176}{780404625}-\frac{22515319 \pi ^2}{2365440}\right) v^6-\frac{16343148032
   v^7}{70945875}+\left(-\frac{169791059264}{869593725}-\frac{33021283 \pi ^2}{2562560}\right) v^8
    \nn\\ &
   -\frac{584895938048
   v^9}{1820944125}+\left(-\frac{53344474395584}{517408266375}-\frac{7894087273 \pi ^2}{492011520}\right)
   v^{10}
  +\cO(v^{11})   \,,
   \end{align}
and
\begin{align}
&\frac{r_{2}(\gamma) -r_{3}(\gamma)}{\pi}=
-\frac{53  }{3 v^3}+\frac{72997  }{5040 v}+\frac{1491  }{400}-\frac{1509 \pi ^2}{140}+\frac{20211 
   v}{640}+\left(\frac{75661 \pi ^2}{4480}-\frac{2678867}{16800}\right) v^2
    \nn\\ &
   +\left(\frac{41053  
   \log\left(\frac{v}{2}\right)}{2450}-\frac{503 \pi ^2}{70}-\frac{123069432361}{4346496000}\right) v^3 +\left(\frac{6139957 
   \pi^2}{394240}-\frac{15259259693}{124185600}\right) v^4
    \nn\\ &
   + \left(\frac{223443793\log
   \left(\frac{v}{2}\right)}{15523200}-\frac{69203 \pi ^2}{6720}+\frac{64307545227137
   }{22375761408000}\right)v^5 +\left(\frac{252585041 \pi ^2}{5857280}-\frac{16101198460801}{56504448000}\right) v^6
    \nn\\ &
   +
   \left(\frac{313945836331 \log \left(\frac{v}{2}\right)}{8879270400}-\frac{346561 \pi ^2}{16896}-\frac{204989896406483131
   }{1828419360768000}\right) v^7 +\left(\frac{918930349 \pi ^2}{11714560}-\frac{2985881877364537 }{8524099584000}\right)
   v^8
    \nn\\ &
   + \left(\frac{16292858440697\log \left(\frac{v}{2}\right)}{307814707200}-\frac{288050119 
   \pi^2}{8785920}-\frac{431015900890630345739 }{1676186519175168000}\right)v^9
    \nn\\ &
   +\left(\frac{326238718783 
   \pi^2}{2788065280}-\frac{98473341190358572667 }{217581903931392000}\right) v^{10}
    +\cO(v^{11}) \,.
\end{align}   
As discussed in Methods, the first six terms of each series were checked against known PN data 
the PN series of $E_{\rm rad}$ and the $\mathbf{\hat b}$ and $\mathbf{p}_{\rm in}$ components of $\mathbf{P}_{\rm recoil}$.
We provide the PN expansion of the impulse in the velocity direction explicitly up to $\mathcal{O}(v^{500})$ in the repository submission.

\section{Graviton Gauge Fixing}\label{sec:gaugefixing}

In order to gauge fix the bulk graviton action, we begin by rewriting the
Einstein-Hilbert term of Eq.~(1) via a total derivative and augment it with
a gauge-fixing term: 
\begin{align}\label{gfactionbulk}
 \mathcal{S}_{\text{bulk}} &= -\frac{1}{16\pi G} \int \mathrm d^{4}x \Big [
 \sqrt{-g} g^{\alpha\beta} g^{\gamma\delta} g^{\mu\nu} ( \Gamma_{\alpha\gamma\delta}
 \Gamma_{\mu\beta\nu} - \Gamma_{\alpha\gamma\mu}  \Gamma_{\delta\beta\nu} )
 +\sfrac{1}{2} G^{\mu}G^{\nu}\eta_{\mu\nu} \Bigr ] \,.
\end{align}
We employ a non-linearly extended de Donder gauge using the gauge fixing vector $G^{\mu}$, the graviton field $h_{\mu\nu}$ (absorbing the coupling $\sqrt{32 \pi G}$), and its trace $h=h_{\mu}{}^{\mu}$ of the form
\begin{align}
\begin{aligned}
G^{\mu} = & \, \partial_{\nu_{1}}h^{\mu\nu_{1}} - \sfrac{1}{2} \partial^{\mu}h 
+\sfrac{1}{2} \partial_{\nu_1}h\,  h^{\mu\nu_{1}}
+\sfrac{1}{2} \partial^{\mu}h_{\nu_1\nu_2}\, h^{\nu_1\nu_2} 
-\partial^{\nu_1}h^{\mu \nu_2} h_{\nu_1\nu_2} 
-\sfrac{1}{4}  \partial^{\mu}h\, h \\ &
+ \partial_{\nu_1}h^{\nu_3\mu}   h^{\nu _1\nu_2}\, h_{\nu_2\nu _3} 
+\sfrac{1}{2} \partial_{\nu_1} h^{\nu_1\nu _2}\, h_{\nu_2\nu_3}\, h^{\nu_3\mu}
-\sfrac{1}{4} \partial_{\nu_2} h\, h_{\nu_2\nu_3} \, h^{\nu_3\mu}
+\partial_{\nu_1}h_{\nu_2\nu_3}\,  h^{\nu_1\nu_2} h^{\nu_3\mu} \\ &
-\sfrac{1}{2} \partial^{\nu_2} h_{\nu_1\nu_2}\,  h^{\mu \nu _1}\, h
+\sfrac{1}{4} \partial^{\mu}h_{\nu_1\nu_2}\, h^{\nu_1\nu_2} h
-\sfrac{1}{2} \partial^{\nu_1} h^{\mu\nu_2}\, h_{\nu_1\nu_2} h
+\sfrac{1}{8} \partial_{\nu_1}h^{\mu\nu_1}\, h^{2}\, .
\end{aligned}
\end{align}
The quadratic-in-$h_{\mu\nu}$ term arising in \eqn{gfactionbulk} yields the graviton propagator of Fig.~3,
whereas the non-linear gauge fixing terms are engineered in such a fashion as 
to maximally simplify the cubic and quartic graviton vertices. The resulting Feynman
vertices up to six-graviton legs along with the used worldline vertices are provided in the 
{\tt feynman rules.m} file.

\section{Expansion of Differential Equation and Single Poles}

In the Methods section, we have constructed a canonical basis satisfying an $\epsilon$-factorized differential equation:
\be
      \frac{\d}{\d x} \vec{J}(x;\epsilon) = \epsilon \hat{A}(x)\, \vec{J}(x;\epsilon) \,.
\label{dgl}
\ee
Besides the factorization of $\epsilon$ in the differential equation \eqref{dgl}, an important property of a canonical basis is that \eqref{dgl} has only simple poles. This guarantees that we only have logarithmic singularities after integration. In other words, when we expand the matrix $\hat{A}(x)$ around all its singularities, i.e., ~for $x=0,1,\infty$, we find at most simple poles.
To test this property, we have to take into account the series expansions of the new
transcendental functions $\varK, \varpi_0, \varpi_1, \alpha_1$, and $G_i$ ---
defined only by the differential equations they satisfy, and thus leaving an ambiguity in their choice.
For this discussion, it is sufficient to study the differential equation \eqref{dgl} around $x=0$,
where we have derived the canonical form, and $x=1$,
around which we solve the differential equation to compute the velocity expansion.
Around both singularities, we provide our local choices of new transcendental functions ---
not the proper analytic continuations of these objects.
In this way, we guarantee that we only exhibit single poles and that all entries of $\hat A(x)$
have a sensible series expansion so that corresponding iterated integrals are easily computed.

The series expansions of our functions are as follows.
Around $x=0$, the CY3 has a special property known as the point of maximal unipotent monodromy. At this point, the basis of periods shows a hierarchy in the number of appearing logarithms, i.e., $\varpi_i^{[0]} \sim \log^i(x)$ for $i=0,1,2,3$. From this structure, it is quite natural to choose $\varpi_0^{[0]},\varpi_1^{[0]}, \alpha_1^{[0]}$ in the following way:
\begin{equation}
\begin{aligned}
	\varpi_0^{[0]} 	&= x+\frac{x^5}{16}+\frac{81 x^9}{4096}+... \, ,\\
	\varpi_1^{[0]} 	&= \log (x) \left(x+\frac{x^5}{16}+\frac{81 x^9}{4096}+...\right)  \, +\frac{x^5}{16}+\frac{189 x^9}{8192}+... \, ,\\
	\alpha_1^{[0]}	&= 1-\frac{x^4}{4}-\frac{93 x^8}{1024}+... \, .
\end{aligned}
\end{equation}
Around $x=1$, the solution space of periods has a different structure containing three power series solutions and just a single logarithmic one. We pick $\varpi_0^{[1]},\varpi_1^{[1]}$ as a linear combination of these three power series solutions: 
\begin{equation}
\begin{aligned}
	\varpi_0^{[1]} 	&= 1-(1-x)+\dfrac{(1-x)^3}{3}+\dfrac{(1-x)^4}{3}+... \, ,\\
	\varpi_1^{[1]} 	&= 1-\frac{(1-x)^2}{2}-\frac{(1-x)^3}{6}+\frac{(1-x)^4}{3}+... \, ,\\
	\alpha_1^{[1]}	&= 1-2\cdot (1-x)+2\cdot(1-x)^2-\dfrac{(1-x)^3}{3}+\dfrac{(1-x)^4}{2}+... \, .
\end{aligned}
\end{equation}
Another (equally valid) choice would change the boundary values of our integrals but not the final physical results.
The K3 surface, however, is special in that it has at both singularities $x=0$ and $x=1$:
a point of maximal unipotent monodromy. Therefore, at both points, there is a natural choice for the period $\varK$. We simply take the unique power series solution, which we can also express through squares of elliptic integrals:
\begin{equation}
\begin{aligned}
	\varK^{[0]} 	&= \dfrac{4}{\pi^2} K^2(x^2) &&=1+\frac{x^2}{2}+\frac{11 x^4}{32}+... \, ,\\
	\varK^{[1]} 	&= \dfrac{4}{\pi^2} K^2(1-x^2) &&=1-(1-x)+\frac{7 (1-x)^2}{8}-\frac{3 (1-x)^3}{4}+... \, .
\end{aligned}
\end{equation}
Given these prerequisites, we may now series expand the differential equation in $(1-x)$:
\be
	\hat{A} 
  =
   \sum_{j=0}^2 \hat{A}_{-1,j} \frac{\log^j(1-x)}{1-x} 
   +
   \sum_{i,j\ge 0}
    \hat{A}_{i,j} (1-x)^i \log^j(1-x) \, .
\label{dglexp}
\ee
We see immediately that there are no higher poles appearing. The terms $(1-x)^{-1}\log^j(1-x)$ for $j=1,2$ are also valid because,
after integration, they give rise only to logarithmic singularities.
These terms arise from the new $G_i$ functions and are consequently associated with the CY3.
This series-expanded form of the differential equation~\eqref{dgl} plays an important role in our matching of boundary integrals
to the slow-velocity $x\to1$ limit.



\section{Boundary matching}


The solution to Eq.~\eqref{dgl} is a path-ordered exponential:
\be
	\vec{J}(x;\epsilon)= \mathcal{P} e^{\epsilon \int_1^x \mathrm dx \hat{A}(x)} \vec{j} \, ,
\label{eq:path_exp}	
\ee
where the vector $\vec{j}$ contains the boundary values of our integrals at $x=1$,
coinciding with the small-velocity limit.
This form can be expanded in $\epsilon$ up to the required order and naturally gives rise
to the iterated integrals defined in Eq.~(4) of the main text.
Our task is now to fix the boundary constants $\vec{j}$ in the small-velocity $x\to1$ limit,
wherein our master integrals become trivial in $x$ but are still non-trivial in $\epsilon$ --- see Section \ref{sec:boundaryintegrals}.
While in principle we could compare every integral with its $x\to1$ limit,
we seek to minimize the number of boundary integrals that must be performed explicitly.
Accordingly, we employ a strategy of imposing analyticity constraints on the general solution of Eq.~\eqref{dgl} ---
relating various boundary values of the master integrals with each other,
and leaving only a small number of unfixed boundary constants.
This is done by also solving the differential equation~\eqref{dgl} perturbatively in $(1-x)$,
while retaining its all-order dependence on $\eps$:
\begin{align}
	\vec J(x;\epsilon) = \sum_{k,m,n}  \vec{f}_{k,m,n}(\epsilon)(1-x)^{k+m\epsilon}\log^{n}(1-x) \,.
\label{eq:wasow}	
\end{align}
The boundary constants are now contained within the functions $\vec{f}_{k,m,n}(\epsilon)$,
and we can easily go from \eqref{eq:wasow} to \eqref{eq:path_exp} by expanding in $\epsilon$.


The $\eps$-resummed solution~\eqref{eq:wasow} is derived from the velocity-expanded form of the differential equation~\eqref{dglexp},
following Wasow's method~\cite{WASOW1966378}.
First, we compute the leading order in $(1-x)$,
i.e.~we set $k=0$ and keep only the terms $\hat A_{-1,i}$ for $i=0,1,2$ in Eq.~\eqref{dglexp}.
In this step, we also determine the range of $m$ in \eqref{eq:wasow}. We write $\vec J = \hat S_0(x;\epsilon) \cdot \vec j + \mathcal O((1-x))$ such that $\hat S_0(x;\epsilon)$ contains all the $(1-x)^{m \epsilon}$ scalings. Then, $\hat S_0$ is determined by:
\be
	\dfrac{\mathrm d \hat{S}_0(x;\epsilon)}{\mathrm dx}=\epsilon \sum_{j=0}^2 \hat{A}_{-1,j}\dfrac{ \log^j{(1-x)}}{1-x}\,  \hat{S}_0(x;\epsilon) \, .
\label{dgl2}	
\ee
Only the entries of $\hat A(x)$ that describe couplings to the CY3 sector contain powers of $\log(1-x)$ in their expansions.
To solve the differential equation~\eqref{dgl2} and compute $\hat{S}_0$, we therefore split our integrals into three groups:
(i) those that do not couple to the CY3 sector (in the top-left corner of Fig.~5, above the Calabi-Yau integrals);
(ii) integrals in the CY3 sector;
(iii) all other integrals coupling through the differential equation to the CY3 sector.
For the first group, $\hat{A}_{-1,1}=\hat{A}_{-1,2}=0$
and so Eq.~\eqref{dgl2} is solved by the matrix exponential $(1-x)^{\epsilon \hat{A}_{-1,0}}$.\footnote{To
compute the matrix exponential quickly, we use the \texttt{Mathematica} package \texttt{Libra}~\cite{Lee:2020zfb},
which makes use of the block-triangular structure of our differential equation.}
For the CY3 sector, we solve \eqref{dgl2} explicitly using~\texttt{Mathematica}'s \texttt{DSolve} command.
For the third group, their homogeneous part is once again given by a matrix exponential since $\hat{A}_{-1,1},\hat{A}_{-1,2}$ are all vanishing.
For the inhomogeneous part, we use the method of variation of constants.
In this way, we have determined $\hat{S}_0$:
the possible $(1-x)$ scalings from Eq.~\eqref{eq:wasow} are $m \in \{10 ,8 ,2,0,-2  ,-4,-6 \}$.

We now continue and compute more terms in the $(1-x)$ expansion of Eq.~\eqref{eq:wasow}.
The summation range of $m$ is completely fixed by the leading order in $(1-x)$,
whereas new powers of $\log(1-x)$ can appear at higher orders in the $(1-x)$ expansion.
Therefore, we rewrite the solution \eqref{eq:wasow},
splitting the previously derived leading-order contributions from higher-order terms:
\be
      \vec{J}= \sum_{i,j}^N \left(\hat{F}_{i,j}(\epsilon) (1-x)^i \log^j{(1-x)}\right) \cdot \hat{S}_0(x;\epsilon)\cdot \vec{j}+\mathcal{O}\left((1-x)^{N+1}\right) \, .
\label{eq:JbigF}
\ee
Here, $N$ is the order in which we want to expand our integrals,
and we have recollected the constants $\vec{f}_{k,m,n}(\epsilon)$ into the matrix $\hat{F}_{i,j}(\epsilon)$ multiplying $\hat{S}_0(x;\epsilon)$.
In this way, we calculate the $\epsilon$-resummed form of our solution to arbitrary order in $(1-x)$. 

Having now determined the slow-velocity $\eps$-resummed solution, we can begin imposing analyticity constraints to fix boundary values.
A general feature of Feynman integrals is that all terms of the form $(1-x)^{m\epsilon}$ with $m>0$ drop out.
Intuitively, this is because the only possible origin of such terms is the integration measure $\mathrm d^{4-2\epsilon} \ell$
of our integrals --- see Fig. 4 of the main text.
Since the measure contains only negative powers of $\epsilon$,
our integrals cannot have positive $\epsilon$ eigenvalues appearing in $\hat{S}_0(x;\epsilon)$ ---
a more rigorous justification is given in refs.~\cite{Henn:2020lye,Dlapa:2023hsl}.
Next, the method of regions yields more precise statements about the $(1-x)^{m\epsilon}$ terms
appearing in the expansions of the initial integrals.
Transforming back to the original (non-canonicalized) basis:
\be
	\lim_{x \rightarrow 1}\vec{I}(x;\epsilon) = \lim_{x \rightarrow 1} \hat T(x;\epsilon)^{-1} \vec{J}(x;\epsilon) \, .
\label{mor}	
\ee
Since the rotation $\hat T(x;\epsilon)^{-1}$ depends non-trivially on $(1-x)$,
we expand Eq.~\eqref{eq:JbigF} to higher orders in $(1-x)$.
Each region comes with a different overall velocity scaling $(1-x)^{-2m\epsilon}$,
where $m$ is the number of radiative gravitons --- scaling as $\ell_i^\text{R}=(\ell_i^0,\ell_i)\sim(v,v)$.
By requiring that the wrong scalings on each side of Eq.~\eqref{mor} cancel, we obtain further relations between the boundary constants. For example, in the (PPP) region, all constants coupling to terms with velocity scaling $m\neq0$ are set to zero. This fixes the majority of our boundary constants.

All remaining boundary constants must now be fixed by comparison with explicit boundary integrals.
On these we perform IBP reduction, 
further reducing the number of manual computations that must be performed.
The remaining boundary integrals that we needed to compute are detailed in Section~\ref{sec:boundaryintegrals}.

\section{Boundary Integrals}\label{sec:boundaryintegrals}

Here we describe our strategy for calculating the boundary integrals required to fix the solutions to the differential equations.
We applied a number of different methods depending on the boundary integral in question.

\subsection{Nomenclature}
Due to the static $v\to 0$ limit, almost all boundary integrals are ladder-type integrals. 
We now introduce a nomenclature to specify every boundary integral in the planar 1SF sector.
Here, three facts about each loop in the planar graphs may be specified:

\begin{enumerate}
\item The type of propagator above the loop. This is typically a worldline $\omega$ or a graviton bubble $g$, but it can also be more sophisticated, such as a “tail-of-tail” bubble.
\item The direction of the propagator, which we specify as a subscript to a propagator. This could be Feynman (nothing), retarded ($+$), advanced ($-$), or a $-i$ times a cut propagator ($\times$).
\item The number of ``jumps", as a subscript to $M$. For brevity, this is omitted when there are no jumps.
\end{enumerate}

\begin{figure}[!h] 
\caption{The $M$, $M_1$, and $M_2$ families, with zero, one, and two jumps, respectively.}
  \begin{tikzpicture}
  \coordinate (inA) at (0.2,.6);
    \coordinate (outA) at (6.2,.6);
    \coordinate (inB) at (0.2,-.6);
    \coordinate (outB) at (6.2,-.6);
    \coordinate (1A) at (.8,.6);
    \coordinate (2A) at (2,.6);
    \coordinate (3A) at (3.2,.6);
    \coordinate (4A) at (4.4,.6);
    \coordinate (5A) at (5.6,.6);
    \coordinate (1B) at (.8,-.6);
    \coordinate (2B) at (2,-.6);
    \coordinate (3B) at (3.2,-.6);
    \coordinate (4B) at (4.4,-.6);
    \coordinate (5B) at (5.6,-.6);
    \draw [fill] (1A) circle (.08);
    \draw [fill] (2A) circle (.08);
    \draw [fill] (3A) circle (.08);
    \draw [fill] (4A) circle (.08);
    \draw [fill] (5A) circle (.08);
    \draw [fill] (1B) circle (.08);
    \draw [fill] (2B) circle (.08);
    \draw [fill] (3B) circle (.08);
    \draw [fill] (4B) circle (.08);
    \draw [fill] (5B) circle (.08);

     \draw [dotted] (inA) -- (outA);
     \draw [dotted] (inB) -- (outB);
     \draw [photon] (1A) -- (1B);
     \draw [photon] (2A) -- (2B);
     \draw [photon] (3A) -- (3B);
     \draw [photon] (4A) -- (4B);
     \draw [photon] (5A) -- (5B); 

     \draw [zUndirected] (1A) -- (2A);
     \draw [zUndirected] (2A) -- (3A);
     \draw [zUndirected] (3A) -- (4A);
     \draw [zUndirected] (4A) -- (5A);
\end{tikzpicture}
  \begin{tikzpicture}
  \coordinate (inA) at (0.2,.6);
    \coordinate (outA) at (6.2,.6);
    \coordinate (inB) at (0.2,-.6);
    \coordinate (outB) at (6.2,-.6);
    \coordinate (1A) at (.8,.6);
    \coordinate (2A) at (2,.6);
    \coordinate (3A) at (3.2,.6);
    \coordinate (4A) at (4.4,.6);
    \coordinate (5A) at (5.6,.6);
    \coordinate (1B) at (.8,-.6);
    \coordinate (2B) at (2,-.6);
    \coordinate (3B) at (3.2,-.6);
    \coordinate (4B) at (4.4,-.6);
    \coordinate (5B) at (5.6,-.6);
    \draw [fill] (1A) circle (.08);
    \draw [fill] (2A) circle (.08);
    \draw [fill] (3A) circle (.08);
    \draw [fill] (4A) circle (.08);
    \draw [fill] (5A) circle (.08);
    \draw [fill] (1B) circle (.08);
    \draw [fill] (2B) circle (.08);
    \draw [fill] (3B) circle (.08);
    \draw [fill] (4B) circle (.08);
    \draw [fill] (5B) circle (.08);

     \draw [dotted] (inA) -- (outA);
     \draw [dotted] (inB) -- (outB);
     \draw [photon] (1A) -- (1B);
     \draw [photon] (2A) -- (2B);
     \draw [photon] (3A) -- (3B);
     \draw [photon] (4A) -- (4B);
     \draw [photon] (5A) -- (5B); 

     \draw [zUndirected] (1A) to[out=45, in=135] (4A);
     \draw [zUndirected] (2A) -- (3A);
     \draw [zUndirected] (3A) -- (4A);
     \draw [zUndirected] (4A) -- (5A);
\end{tikzpicture}
  \begin{tikzpicture}
  \coordinate (inA) at (0.2,.6);
    \coordinate (outA) at (6.2,.6);
    \coordinate (inB) at (0.2,-.6);
    \coordinate (outB) at (6.2,-.6);
    \coordinate (1A) at (.8,.6);
    \coordinate (2A) at (2,.6);
    \coordinate (3A) at (3.2,.6);
    \coordinate (4A) at (4.4,.6);
    \coordinate (5A) at (5.6,.6);
    \coordinate (1B) at (.8,-.6);
    \coordinate (2B) at (2,-.6);
    \coordinate (3B) at (3.2,-.6);
    \coordinate (4B) at (4.4,-.6);
    \coordinate (5B) at (5.6,-.6);
    \draw [fill] (1A) circle (.08);
    \draw [fill] (2A) circle (.08);
    \draw [fill] (3A) circle (.08);
    \draw [fill] (4A) circle (.08);
    \draw [fill] (5A) circle (.08);
    \draw [fill] (1B) circle (.08);
    \draw [fill] (2B) circle (.08);
    \draw [fill] (3B) circle (.08);
    \draw [fill] (4B) circle (.08);
    \draw [fill] (5B) circle (.08);

     \draw [dotted] (inA) -- (outA);
     \draw [dotted] (inB) -- (outB);
     \draw [photon] (1A) -- (1B);
     \draw [photon] (2A) -- (2B);
     \draw [photon] (3A) -- (3B);
     \draw [photon] (4A) -- (4B);
     \draw [photon] (5A) -- (5B); 

     \draw [zUndirected] (1A) to[out=45, in=135] (4A);
     \draw [zUndirected] (2A) to[out=45, in=135] (4A);
     \draw [zUndirected] (3A) -- (4A);
     \draw [zUndirected] (4A) -- (5A);
\end{tikzpicture}
\end{figure}
In this notation, the cut identity
\be
-i\dd(x) = \frac1{x+\ii0^+} - \frac1{x-\ii0^+} 
\ee
can be re-expressed as
\be
M[\ldots,\omega_\times,\ldots] = M[\ldots,\omega_+,\ldots] - M[\ldots,\omega_-,\ldots] \, .
\ee
In the general case, we must calculate all possible $\ii0^+$ routings. Therefore, given an integral calculated with one $\ii0^+$ routing and the cut, which factorizes into lower-loop integrals, we obtain the opposite $\ii0^+$ routing ``for free".

To clarify the nomenclature used in the repository, we give some specific examples. Our convention for the diagrams is that arrows pointing right indicate a retarded $1/(\ell \cdot v + \ii0^+)$ propagator, while arrows pointing left also indicate a retarded propagator, but with opposite momentum flow $1/(-\ell \cdot v + \ii0^+)$.
\begin{align}
M[g_+ g_+ g_-] &= \raisebox{-0.6cm}{
\begin{tikzpicture}
  \coordinate (inA) at (0.2,.6);
    \coordinate (outA) at (5,.6);
    \coordinate (inB) at (0.2,-.6);
    \coordinate (outB) at (5,-.6);
    \coordinate (1A) at (.8,.6);
    \coordinate (2A) at (2,.6);
    \coordinate (3A) at (3.2,.6);
    \coordinate (4A) at (4.4,.6);
    \coordinate (1B) at (.8,-.6);
    \coordinate (2B) at (2,-.6);
    \coordinate (3B) at (3.2,-.6);
    \coordinate (4B) at (4.4,-.6);
    \coordinate (23C) at (2.6, .8);
    \draw [fill] (1A) circle (.08);
    \draw [fill] (2A) circle (.08);
    \draw [fill] (3A) circle (.08);
    \draw [fill] (4A) circle (.08);
    \draw [fill] (1B) circle (.08);
    \draw [fill] (4B) circle (.08);
     \draw [dotted] (inA) -- (outA);
     \draw [dotted] (inB) -- (outB);
     \draw [photon] (1A) -- (1B);
     \draw [photon] (4A) -- (4B);
     \draw [photon] (1A) to[out=80, in=100] (2A);
     \draw [zParticle] (0.8,0.9) to[out=60, in=120] (2,0.9);
     \draw [photon] (2A) to[out=80, in=100] (3A);
     \draw [zParticle] (2,0.9) to[out=60, in=120] (3.2,0.9);
     \draw [photon] (3A) to[out=80, in=100] (4A);
     \draw [zParticle] (4.4,0.9) to[out=120, in=60] (3.2,0.9);
\end{tikzpicture}}
&&\hspace{-3ex}= -\frac{1}{12288 \pi ^4} + \frac{ 12 \gamma_E -41-6\log (2)-12 \log (\pi )}{36864 \pi ^4} \epsilon + \mathcal{O}(\epsilon^2) \, ,\\
M[0,g_+,0] &= \raisebox{-0.6cm}{
\begin{tikzpicture}
  \coordinate (inA) at (0.6,.6);
    \coordinate (outA) at (4.6,.6);
    \coordinate (inB) at (0.6,-.6);
    \coordinate (outB) at (4.6,-.6);
    \coordinate (2A) at (1.8,.6);
    \coordinate (3A) at (3.4,.6);
    \coordinate (1B) at (1.2,-.6);
    \coordinate (2B) at (2.4,-.6);
    \coordinate (3B) at (2.8,-.6);
    \coordinate (4B) at (4.0,-.6);
    \coordinate (23C) at (2.6, .8);
    \draw [fill] (2A) circle (.08);
    \draw [fill] (3A) circle (.08);
    \draw [fill] (1B) circle (.08);
    \draw [fill] (2B) circle (.08);
    \draw [fill] (3B) circle (.08);
    \draw [fill] (4B) circle (.08);
     \draw [dotted] (inA) -- (outA);
     \draw [dotted] (inB) -- (outB);
     \draw [photon] (2A) -- (1B);
     \draw [photon] (2A) -- (2B);
     \draw [photon] (3A) -- (3B);
     \draw [photon] (3A) -- (4B);
     \draw [photon] (2A) to[out=80, in=100] (3A);
     \draw [zParticle] (1.8,1.0) to[out=50, in=130] (3.4,1.0);
\end{tikzpicture}}
&&\hspace{-3ex}=  -\frac{1}{65536 \pi ^2} -\frac{  -4 \gamma_E +3+26 \log (2)+4 \log (\pi )}{65536 \pi ^2} \epsilon + \mathcal{O}(\epsilon^2) \, ,\\
M[\omega_+,g_+,\omega_-] &=   -\hspace{-1ex}\raisebox{-0.6cm}{
\begin{tikzpicture}
  \coordinate (inA) at (0.2,.6);
    \coordinate (outA) at (5,.6);
    \coordinate (inB) at (0.2,-.6);
    \coordinate (outB) at (5,-.6);
    \coordinate (1A) at (.8,.6);
    \coordinate (2A) at (2,.6);
    \coordinate (3A) at (3.2,.6);
    \coordinate (4A) at (4.4,.6);
    \coordinate (1B) at (.8,-.6);
    \coordinate (2B) at (2,-.6);
    \coordinate (3B) at (3.2,-.6);
    \coordinate (4B) at (4.4,-.6);
    \coordinate (23C) at (2.6, .8);
    \draw [fill] (1A) circle (.08);
    \draw [fill] (2A) circle (.08);
    \draw [fill] (3A) circle (.08);
    \draw [fill] (4A) circle (.08);
    \draw [fill] (1B) circle (.08);
    \draw [fill] (2B) circle (.08);
    \draw [fill] (3B) circle (.08);
    \draw [fill] (4B) circle (.08);
     \draw [dotted] (inA) -- (outA);
     \draw [dotted] (inB) -- (outB);
     \draw [photon] (1A) -- (1B);
     \draw [photon] (2A) -- (2B);
     \draw [photon] (3A) -- (3B);
     \draw [photon] (4A) -- (4B);
     \draw [zParticle] (1A) -- (2A);
     \draw [photon] (2A) to[out=80, in=100] (3A);
     \draw [zParticle] (4A) -- (3A); 
     \draw [zParticle] (2,0.9) to[out=60, in=120] (3.2,0.9);
\end{tikzpicture}} 
&&\hspace{-3ex}= \frac{1}{8192 \pi ^4 \epsilon ^2}-\frac{2 \gamma_E -3-3\log (2 ) - 2 \log(\pi)}{4096 \pi ^4 \epsilon } + \mathcal{O}(\epsilon^0) \, .
\end{align}
Note the relative sign in the third line due to the conversion of $\omega_-$ to a left-pointing retarded propagator.

\subsection{General Formulae}

In the loop-by-loop approach, we principally make use of the following integral, where $\int_\ell=\int \frac{\mathrm d^D \ell}{(2 \pi)^D}$:
\begin{align}\label{eq:susy}
\int_\ell\frac{\dd(\ell\cdot v_1)}{(\ell^2+\ii0)^{\nu_1}((\ell-q)^2+\ii0)^{\nu_2}(\ell\cdot v_2+\ii0)^{\nu_3}}=&(4\pi)^{\frac{1-D}{2}}(-1)^{\nu_1+\nu_2+\nu_3}i^{\nu_3}|q|^{D-1-2\nu_1-2\nu_2-\nu_3}\\ &\times\frac{\Gamma(\nu_1+\nu_2+\frac{\nu_3}{2}-\frac{D-1}{2})\Gamma(\frac{\nu_3}{2})\Gamma(\frac{D-1}{2}-\nu_1-\frac{\nu_3}{2})\Gamma(\frac{D-1}{2}-\nu_2-\frac{\nu_3}{2})}{2 \Gamma(\nu_1)\Gamma(\nu_2)\Gamma(\nu_3)\Gamma(D-1-\nu_1-\nu_2-\nu_3)}\,,\nn
\end{align}
where $q \cdot v_1 = q \cdot v_2=0$. In practice, this means that a loop-by-loop integration over linear propagators is only possible when a linear propagator is next to a cut.
When $\nu_3=0$, no such restriction exists, and we use
\be \label{eq:empty_loop}
\int_\ell\frac{\dd(\ell\cdot v_1)}{(\ell^2+\ii0)^{\nu_1}((\ell-q)^2+\ii0)^{\nu_2}}=(4\pi)^{\frac{1-D}{2}}(-1)^{\nu_1+\nu_2}|q|^{D-1-2\nu_1-2\nu_2} 
\frac{\Gamma \left(\nu_1+\nu_2-\frac{D-1}{2}\right) \Gamma \left(\frac{D-1}{2}-\nu_1\right) \Gamma \left(\frac{D-1}{2}-\nu_2\right)}{\Gamma \left(\nu _1\right) \Gamma \left(\nu _2\right) \Gamma \left(D-1 -\nu _1-\nu _2\right)} \, .
\ee 
For example, $M[\omega_+]$ is a special case of Eq.~\eqref{eq:susy} where $\nu_1 = \nu_2 = \nu_3 = 1$
(each delta function comes with a factor $2 \pi$ and each loop integral comes with a factor $1/(2\pi)^4$):
\begin{align}
M[\omega_+] = 
\vcenter{\hbox{\begin{tikzpicture}
  \coordinate (inA) at (0.2,.6);
    \coordinate (outA) at (2.6,.6);
    \coordinate (inB) at (0.2,-.6);
    \coordinate (outB) at (2.6,-.6);
    \coordinate (1A) at (.8,.6);
    \coordinate (2A) at (2,.6);
    \coordinate (1B) at (.8,-.6);
    \coordinate (2B) at (2,-.6);
    \draw [fill] (1A) circle (.08);
    \draw [fill] (2A) circle (.08);
    \draw [fill] (1B) circle (.08);
    \draw [fill] (2B) circle (.08);
     \draw [dotted] (inA) -- (outA);
     \draw [dotted] (inB) -- (outB);
     \draw [photon] (1A) -- (1B);
     \draw [photon] (2A) -- (2B);
     \draw [zParticle] (1A) -- (2A);
\end{tikzpicture}}}
 = -i (4 \pi )^{\epsilon -1} |q| ^{-2 (\epsilon +1)} \frac{\Gamma (-\epsilon )^2 \Gamma
   (\epsilon +1)}{2 \Gamma (-2 \epsilon )}.
\end{align}
Furthermore, in the radiative regions, we encounter ``graviton bubbles", which can be integrated out:
\begin{align}
  \vcenter{\hbox{\begin{tikzpicture}
  \coordinate (inA) at (0.2,.6);
    \coordinate (outA) at (2.6,.6);
    \coordinate (inB) at (0.2,-.6);
    \coordinate (outB) at (2.6,-.6);
    \coordinate (1A) at (.8,.6);
    \coordinate (2A) at (2,.6);
    \coordinate (1B) at (.8,-.6);
    \coordinate (2B) at (2,-.6);
    \draw [fill] (1A) circle (.08);
    \draw [fill] (2A) circle (.08);
    \draw [fill] (1B) circle (.08);
    \draw [fill] (2B) circle (.08);
     \draw [dotted] (inA) -- (outA);
     \draw [dotted] (inB) -- (outB);
     \draw [photon] (1A) -- (1B);
     \draw [photon] (2A) -- (2B);
     \draw [photon] (1A) to[out=80, in=100] (2A);
     \draw [zParticle] (0.8,0.9) to[out=60, in=120] (2,0.9);
\end{tikzpicture}}}
\rightarrow
  \vcenter{\hbox{\begin{tikzpicture}
  \coordinate (inA) at (0.2,.6);
    \coordinate (outA) at (2.6,.6);
    \coordinate (inB) at (0.2,-.6);
    \coordinate (outB) at (2.6,-.6);
    \coordinate (1A) at (.8,.6);
    \coordinate (2A) at (2,.6);
    \coordinate (1B) at (.8,-.6);
    \coordinate (2B) at (2,-.6);
    \draw [fill] (1A) circle (.08);
    \draw [fill] (2A) circle (.08);
    \draw [fill] (1B) circle (.08);
    \draw [fill] (2B) circle (.08);
     \draw [dotted] (inA) -- (outA);
     \draw [dotted] (inB) -- (outB);
     \draw [photon] (1A) -- (1B);
     \draw [photon] (2A) -- (2B);
     \draw [zParticleBold] (1A) -- (2A);
     \draw (1.9, 0.6) node [above] {$(\ell \cdot v_1)^{(-1+2\epsilon)}$};
\end{tikzpicture}}}
\label{fig:integratingOutBubble}
\end{align}
This leads to a linear propagator with an $\eps$-dependent, non-integer power:
\be \label{eq:graviton_bubble}
\int_\ell\frac{\dd(\ell\cdot v)}{(\ell-k)^2+\ii0^+\left(k-\ell\right)\cdot v}=-(+i)^{1-2\epsilon}(k\cdot v+\ii0)^{1-2\epsilon} (4\pi)^{\frac{2\epsilon-3}2}\Gamma\left(\frac{2\epsilon-1}2\right)\, ,
\ee

\subsection{Schwinger parametrization}

Integrals that cannot be determined using loop-by-loop integration, or fully constrained via cuts and partial fractions,
we obtain using Schwinger parametrization. We employ the following identities:
\be
\begin{aligned}
  \frac{1}{(\omega+\ii0^+)^\alpha}
  &=
  e^{-i\frac{\pi}{2}\alpha}\frac{1}{\Gamma(\alpha)}\int_0^\infty \mathrm du\, u^{\alpha-1}e^{-u(0-i\omega)} \, ,
  \\
\frac{1}{(\omega-\ii0^+)^\alpha}
&=
e^{i\frac{\pi}{2}\alpha}\frac{1}{\Gamma(\alpha)}\int_0^\infty \mathrm du \, u^{\alpha-1}e^{-u(0+i\omega)} \, , 
\\
\frac{1}{(\boldsymbol{\ell}^2)^\alpha} 
&=
 \frac{1}{\Gamma(\alpha)} \int_0^\infty \mathrm du \, u^{\alpha-1} e^{-u\boldsymbol{\ell}^2} \, .
\end{aligned}
\ee
Prior to Schwinger parametrization, we integrate out all loops possible using Eqs.~\eqref{eq:empty_loop} and \eqref{eq:graviton_bubble}. Once the above identities have been applied, integration over loop momenta is trivialized as integrals over Gaussians, leaving behind a non-trivial integral of the Schwinger parameters. There is no systematic procedure for solving these integrals, although we note that \texttt{Mathematica} is more effective when integrating over Schwinger parameters corresponding to linear propagators first.

Where graviton bubbles with opposite $\ii0^+$ prescription appear on the same loop, e.g. $M[g_+\times g_-]$, we apply the following formula (valid for at least one non-integer power), separating the different prescriptions into separate integrals:
\be
\frac1{(\omega+\ii0^+)^\alpha(\omega-\ii0^+)^\beta}=e^{-i\pi\alpha}\frac{\sin[\pi\beta]}{\sin[\pi(\alpha+\beta)]}\frac1{(\omega-\ii0^+)^{\alpha+\beta}}+e^{i\pi\beta}\frac{\sin[\pi\alpha]}{\sin[\pi(\alpha+\beta)]}\frac1{(\omega+\ii0^+)^{\alpha+\beta}} \, .
\ee

A direct corollary of this is that we may convert an arbitrary combination of $\ii0^{+}$ prescriptions in a successive row of graviton jumps and worldline propagators into the sum of two uniform configurations using the formula:

\begin{align}
 \raisebox{-0.35cm}{  \begin{tikzpicture}
  \coordinate (inA) at (0.6,.6);
    \coordinate (outA) at (6.5,.6);
    \coordinate (inB) at (0.6,-.6);
    \coordinate (outB) at (6.5,-.6);
    \coordinate (1A) at (.8,.6);
    \coordinate (2A) at (1.6,.6);
    \coordinate (3A) at (2.4,.6);
    \coordinate (4A) at (3.2,.6);
     \coordinate (5A) at (4.0,.6);
     \coordinate (6A) at (4.8,.6);
     \coordinate (7A) at (5.6,.6);
      \coordinate (8A) at (6.4,.6);
    \coordinate (1B) at (.8,-.6);
    \coordinate (2B) at (1.6,-.6);
    \coordinate (3B) at (2.4,-.6);
    \coordinate (4B) at (3.2,-.6);
    \coordinate (5B) at (4.0, -.6);
     \coordinate (6B) at (4.8, -.6);
     \coordinate (7B) at (5.6,-.6);
     \coordinate (8B) at (6.4,-.6);
    \draw [fill] (1A) circle (.08);
    \draw [fill] (2A) circle (.08);
    \draw [fill] (3A) circle (.08);
    \draw [fill] (4A) circle (.08);
    \draw [fill] (5A) circle (.08);
    \draw [fill] (6A) circle (.08);
    \draw [fill] (7A) circle (.08);
     \draw [fill] (8A) circle (.08);
         \draw [dotted] (2A) --  node [below] {$n_{1}$} (3A);
          \draw [dotted] (4A) -- node [below] {$n_{2}$} (5A);
     \draw [dotted, ultra thick] (6A) -- (7A);
     \draw [zUndirected] (1A) -- node [above] {$s_{1}$} node [below] {$\nu_{1}$} (2A) ;
          \draw [photon] (2A)   to[out=90, in=90] node [above] {$t_{1}$} (3A);
     \draw [zUndirected] (3A) -- node [above] {$s_{2}$} node [below] {$\nu_{2}$} (4A); 
     \draw [photon] (4A) to[out=90, in=90] node [above] {$t_{2}$}  (5A);
       \draw [dotted, ultra thick] (6A) to[out=90, in=90]   (7A);
     \draw [zUndirected] (5A) -- node [above] {$s_{3}$} node [below] {$\nu_{3}$} (6A);  
      \draw [zUndirected] (7A) -- node [above] {$s_{n}$} node [below] {$\nu_{n}$} (8A); 
\end{tikzpicture}}
&=
\frac{\sin(\pi \alpha_{+})}{\sin(\pi[ \alpha_{+}+\alpha_{-}])}
\raisebox{0cm}{  \begin{tikzpicture}
  \coordinate (inA) at (0.6,.6);
    \coordinate (outA) at (6.5,.6);
    \coordinate (inB) at (0.6,-.6);
    \coordinate (outB) at (6.5,-.6);
    \coordinate (1A) at (.8,.6);
    \coordinate (2A) at (1.6,.6);
    \coordinate (3A) at (2.4,.6);
    \coordinate (4A) at (3.2,.6);
     \coordinate (5A) at (4.0,.6);
     \coordinate (6A) at (4.8,.6);
     \coordinate (7A) at (5.6,.6);
     \coordinate (8A) at (6.4,.6);
    \coordinate (1B) at (.8,-.6);
    \coordinate (2B) at (1.6,-.6);
    \coordinate (3B) at (2.4,-.6);
    \coordinate (4B) at (3.2,-.6);
    \coordinate (5B) at (4.0, -.6);
     \coordinate (6B) at (4.8, -.6);
     \coordinate (7B) at (5.6,-.6);
     \coordinate (8B) at (6.4,-.6);
    \draw [fill] (1A) circle (.08);
    \draw [fill] (2A) circle (.08);
    \draw [fill] (3A) circle (.08);
    \draw [fill] (4A) circle (.08);
    \draw [fill] (5A) circle (.08);
    \draw [fill] (6A) circle (.08);
    \draw [fill] (7A) circle (.08);
     \draw [fill] (8A) circle (.08);
     \draw [dotted] (1A) -- (6A);
     \draw [dotted, ultra thick] (6A) -- (7A);
      \draw [zParticle] (7A) -- (8A); 
     \draw [zParticle] (1A) -- (2A);
     \draw [photon] (2A) to[out=90, in=90] (3A);
     \draw [zParticle] (3A) -- (4A); 
     \draw [photon] (4A) to[out=90, in=90] (5A);
     \draw [zParticle] (5A) -- (6A);  
      \draw [dotted, ultra thick] (6A) to[out=90, in=90]   (7A);
      \draw [zParticle] (1.6,0.9) to[out=60, in=120] (2.4,0.9);
       \draw [zParticle] (3.2,0.9) to[out=60, in=120] (4.0,0.9);
\end{tikzpicture}}
\nn\\ &
+\frac{\sin(\pi \alpha_{-})}{\sin(\pi[ \alpha_{+}+\alpha_{-}])}
 \raisebox{0cm}{  \begin{tikzpicture}
  \coordinate (inA) at (0.4,.6);
    \coordinate (outA) at (6.0,.6);
    \coordinate (inB) at (0.4,-.6);
   \coordinate (outB) at (6.0,-.6);
    \coordinate (1A) at (.8,.6);
    \coordinate (2A) at (1.6,.6);
    \coordinate (3A) at (2.4,.6);
    \coordinate (4A) at (3.2,.6);
     \coordinate (5A) at (4.0,.6);
     \coordinate (6A) at (4.8,.6);
     \coordinate (7A) at (5.6,.6);
    \coordinate (1B) at (.8,-.6);
    \coordinate (2B) at (1.6,-.6);
    \coordinate (3B) at (2.4,-.6);
    \coordinate (4B) at (3.2,-.6);
    \coordinate (5B) at (4.0, -.6);
     \coordinate (6B) at (4.8, -.6);
     \coordinate (7B) at (5.6,-.6);
    \draw [fill] (1A) circle (.08);
    \draw [fill] (2A) circle (.08);
    \draw [fill] (3A) circle (.08);
    \draw [fill] (4A) circle (.08);
    \draw [fill] (5A) circle (.08);
    \draw [fill] (6A) circle (.08);
    \draw [fill] (7A) circle (.08);
    \draw [fill] (8A) circle (.08);
    \draw [dotted] (1A) -- (6A);
     \draw [dotted, ultra thick] (6A) -- (7A);
     \draw [zParticle] (2A) -- (1A);
     \draw [photon] (3A) to[in=90, out=90] (2A);
     \draw [zParticle] (4A) -- (3A); 
     \draw [photon] (5A) to[out=90, in=90] (4A);
     \draw [zParticle] (6A) -- (5A);  
     \draw [zParticle] (8A) -- (7A);  
      \draw [dotted, ultra thick] (7A) to[out=90, in=90]   (6A);
      \draw [zParticle] (2.4,0.9) to[out=120, in=60] (1.6,0.9);
       \draw [zParticle] (4.0,0.9) to[out=120, in=60] (3.2,0.9);
\end{tikzpicture}} \, ,
\end{align}
where the $s_{i}=\pm \ii0^{+}$ and $t_{i}=\pm \ii0^{+}$ yield the $\ii0^{+}$ prescriptions
on the left-hand-side and ($D=4-2\epsilon$)
\be
\alpha_{\pm}= \sum_{i|s_{i}=\pm \ii0^{+}} \nu_{i} + \sum_{a|t_{a}=\pm \ii0^{+}} (2n_{a}-3 + 2\epsilon)
\, .
\ee
Here, we generalized to powers $\nu_{i}$ for the worldline and $n_{a}$ for the graviton propagators, while in our applications, $\nu_{i}=n_{a}=1$ always.

\subsection{Tail-of-tail integrals}

\begin{figure}[!h]
  \begin{tikzpicture}
  \coordinate (inA) at (0.8,.6);
    \coordinate (outA) at (4.4,.6);
    \coordinate (inB) at (0.8,-.6);
    \coordinate (outB) at (4.4,-.6);
    \coordinate (2A) at (1.4,.6);
    \coordinate (3A) at (3.8,.6);
    \coordinate (2B) at (1.4,-.6);
    \coordinate (3B) at (3.8,-.6);
    \coordinate (23C) at (2.6, 1.2);
    \coordinate (21A) at (2.2, 0.6);
    \coordinate (22A) at (3.0, 0.6);
    \draw [fill] (2A) circle (.08);
    \draw [fill] (3A) circle (.08);
    \draw [fill] (2B) circle (.08);
    \draw [fill] (3B) circle (.08);
    \draw [fill] (23C) circle (.08);
     \draw [fill] (21A) circle (.08);
     \draw [fill] (22A) circle (.08);
    
     \draw [dotted] (inA) -- (outA);
     \draw [dotted] (inB) -- (outB);
     \draw [photon] (2A) -- (2B);
     \draw [photon] (3A) -- (3B);
     
     \draw [photon] (2A) to[out=90, in=180] (23C);
     \draw [photon] (23C) to[out=0, in=90] (3A);
     \draw [photon] (23C) -- (21A);
     \draw [photon] (23C) -- (22A);
     \draw [zParticle] (1.4,1.0) to[out=60, in=180] (2.55, 1.4);
     \draw [zParticle] (2.65, 1.4) to[out=0, in=120] (3.8, 1.0);
     
     \draw (1.9, 0.6) node [above] {$\ell_1$};
     \draw (2.65, 0.6) node [above] {$\ell_2$};
     \draw (3.3, 0.6) node [above] {$\ell_3$};
     \draw (2.65, 0.6) node [below] {$\ell$};
     
\end{tikzpicture}
  \begin{tikzpicture}
  \coordinate (inA) at (0.8,.6);
    \coordinate (outA) at (4.4,.6);
    \coordinate (inB) at (0.8,-.6);
    \coordinate (outB) at (4.4,-.6);
    \coordinate (2A) at (1.4,.6);
    \coordinate (3A) at (3.8,.6);
    \coordinate (2B) at (1.4,-.6);
    \coordinate (3B) at (3.8,-.6);
    \coordinate (23C) at (2.6, 1.2);
    \coordinate (21A) at (2.2, 0.6);
    \coordinate (22A) at (3.0, 0.6);
    \draw [fill] (2A) circle (.08);
    \draw [fill] (3A) circle (.08);
    \draw [fill] (2B) circle (.08);
    \draw [fill] (3B) circle (.08);
    \draw [fill] (23C) circle (.08);
     \draw [fill] (21A) circle (.08);
     \draw [fill] (22A) circle (.08);
    
     \draw [dotted] (inA) -- (outA);
     \draw [dotted] (inB) -- (outB);
     \draw [photon] (2A) -- (2B);
     \draw [photon] (3A) -- (3B);
     
     \draw [photon] (2A) to[out=90, in=180] (23C);
     \draw [photon] (23C) to[out=0, in=90] (3A);
     \draw [photon] (23C) -- (21A);
     \draw [photon] (23C) -- (22A);
     \draw [zParticle] (1.4,1.0) to[out=60, in=180] (2.55, 1.4);
     \draw [zParticle] (3.8, 1.0) to[out=120, in=0] (2.65, 1.4);
\end{tikzpicture}
  \begin{tikzpicture}
    \coordinate (inA) at (0.8,.6);
    \coordinate (outA) at (5.6,.6);
    \coordinate (inB) at (0.8,-.6);
    \coordinate (outB) at (5.6,-.6);
    \coordinate (2A) at (1.4,.6);
    \coordinate (3A) at (3.8,.6);
    \coordinate (4A) at (5.0,.6);
    \coordinate (2B) at (1.4,-.6);
    \coordinate (3B) at (3.8,-.6);
    \coordinate (4B) at (5.0,-.6);
    \coordinate (23C) at (2.6, 1.2);
    \coordinate (21A) at (2.2, 0.6);
    \coordinate (22A) at (3.0, 0.6);
    \draw [fill] (2A) circle (.08);
    \draw [fill] (3A) circle (.08);
    \draw [fill] (4A) circle (.08);
    \draw [fill] (2B) circle (.08);
    \draw [fill] (4B) circle (.08);
    \draw [fill] (23C) circle (.08);
     \draw [fill] (21A) circle (.08);
     \draw [fill] (22A) circle (.08);
    
     \draw [dotted] (inA) -- (outA);
     \draw [dotted] (inB) -- (outB);
     \draw [photon] (2A) -- (2B);
     \draw [photon] (4A) -- (4B);
     
     \draw [zParticle] (3A) -- (4A);
     
     \draw [photon] (2A) to[out=90, in=180] (23C);
     \draw [photon] (23C) to[out=0, in=90] (3A);
     \draw [photon] (23C) -- (21A);
     \draw [photon] (23C) -- (22A);
     \draw [zParticle] (1.4,1.0) to[out=60, in=180] (2.55, 1.4);
     \draw [zParticle] (2.65, 1.4) to[out=0, in=120] (3.8, 1.0);
\end{tikzpicture}
  \begin{tikzpicture}
    \coordinate (inA) at (0.8,.6);
    \coordinate (outA) at (5.6,.6);
    \coordinate (inB) at (0.8,-.6);
    \coordinate (outB) at (5.6,-.6);
    \coordinate (2A) at (1.4,.6);
    \coordinate (3A) at (3.8,.6);
    \coordinate (4A) at (5.0,.6);
    \coordinate (2B) at (1.4,-.6);
    \coordinate (3B) at (3.8,-.6);
    \coordinate (4B) at (5.0,-.6);
    \coordinate (23C) at (2.6, 1.2);
    \coordinate (21A) at (2.2, 0.6);
    \coordinate (22A) at (3.0, 0.6);
    \draw [fill] (2A) circle (.08);
    \draw [fill] (3A) circle (.08);
    \draw [fill] (4A) circle (.08);
    \draw [fill] (2B) circle (.08);
    \draw [fill] (4B) circle (.08);
    \draw [fill] (23C) circle (.08);
     \draw [fill] (21A) circle (.08);
     \draw [fill] (22A) circle (.08);
    
     \draw [dotted] (inA) -- (outA);
     \draw [dotted] (inB) -- (outB);
     \draw [photon] (2A) -- (2B);
     \draw [photon] (4A) -- (4B);
     
     \draw [zParticle] (3A) -- (4A);
     
     \draw [photon] (2A) to[out=90, in=180] (23C);
     \draw [photon] (23C) to[out=0, in=90] (3A);
     \draw [photon] (23C) -- (21A);
     \draw [photon] (23C) -- (22A);
     \draw [zParticle] (1.4,1.0) to[out=60, in=180] (2.55, 1.4);
     \draw [zParticle] (3.8, 1.0) to[out=120, in=0] (2.65, 1.4);
\end{tikzpicture}

\caption{The required tail-of-tail integrals $M[\text{tail}_+\text{tail}_+]$, $M[\text{tail}_+\text{tail}_-]$, $M[\text{tail}_+\text{tail}_+ \omega_+]$, and $M[\text{tail}_+\text{tail}_- \omega_+]$.}\label{fig:tailtail}
\end{figure}

A new class of boundary integrals is given by the integrals shown in Fig.~\ref{fig:tailtail}.
We refer to these as tail-of-tail (boundary) integrals following ref.~\cite{Edison:2023qvg}. In total, we require four of those as shown in Fig.~\ref{fig:tailtail}.
We compute these as iterated integrals.
Exactly as with the simple graviton bubbles, Eq.~\ref{fig:integratingOutBubble}, the more complicated ``tail-of-tail bubble" depends only on a single scale, its energy.
We may therefore integrate this bubble away first resulting in a worldline propagators with an $\eps$-dependent power:
\begin{align}
  \vcenter{\hbox{
  \begin{tikzpicture}
    \coordinate (inA) at (0.8,.6);
      \coordinate (outA) at (4.4,.6);
      \coordinate (inB) at (0.8,-.6);
      \coordinate (outB) at (4.4,-.6);
      \coordinate (2A) at (1.4,.6);
      \coordinate (3A) at (3.8,.6);
      \coordinate (23C) at (2.6, 1.2);
      \coordinate (21A) at (2.2, 0.6);
      \coordinate (22A) at (3.0, 0.6);
      \draw [fill] (2A) circle (.08);
      \draw [fill] (3A) circle (.08);
      \draw [fill] (23C) circle (.08);
       \draw [fill] (21A) circle (.08);
       \draw [fill] (22A) circle (.08);
       \draw [zParticle] (inA) -- (2A);
       \draw [zParticle] (3A) -- (outA);
       \draw [dotted] (3A) -- (2A);
       \draw [photon] (2A) to[out=90, in=180] (23C);
       \draw [photon] (23C) to[out=0, in=90] (3A);
       \draw [photon] (23C) -- (21A);
       \draw [photon] (23C) -- (22A);
       \draw [zParticle] (1.4,1.0) to[out=60, in=180] (2.55, 1.4);
       \draw [zParticle] (2.65, 1.4) to[out=0, in=120] (3.8, 1.0);
      \draw (inA) node [above] {$\omega$} ;
      \draw (outA) node [above] {$\omega$} ;
  \end{tikzpicture}}}
  &=
  f(\eps)
  (-\ii(\omega+\ii \zz))^{1-6\eps}
  \,,
  \\
  \vcenter{\hbox{
    \begin{tikzpicture}
      \coordinate (inA) at (0.8,.6);
        \coordinate (outA) at (4.4,.6);
        \coordinate (inB) at (0.8,-.6);
        \coordinate (outB) at (4.4,-.6);
        \coordinate (2A) at (1.4,.6);
        \coordinate (3A) at (3.8,.6);
        \coordinate (23C) at (2.6, 1.2);
        \coordinate (21A) at (2.2, 0.6);
        \coordinate (22A) at (3.0, 0.6);
        \draw [fill] (2A) circle (.08);
        \draw [fill] (3A) circle (.08);
        \draw [fill] (23C) circle (.08);
         \draw [fill] (21A) circle (.08);
         \draw [fill] (22A) circle (.08);
         \draw [zParticle] (inA) -- (2A);
         \draw [zParticle] (3A) -- (outA);
         \draw [dotted] (3A) -- (2A);
         \draw [photon] (2A) to[out=90, in=180] (23C);
         \draw [photon] (23C) to[out=0, in=90] (3A);
         \draw [photon] (23C) -- (21A);
         \draw [photon] (23C) -- (22A);
         \draw [zParticle] (1.4,1.0) to[out=60, in=180] (2.55, 1.4);
         \draw [zParticle] (3.8, 1.0) to[out=120, in=0] (2.65, 1.4);
        \draw (inA) node [above] {$\omega$} ;
        \draw (outA) node [above] {$\omega$} ;
    \end{tikzpicture}}}
    &=
    f(\eps)
    (\omega+\ii \zz)^{\frac{1-6\eps}{2}}
    (\omega-\ii \zz)^{\frac{1-6\eps}{2}}
    \,.
\end{align}
The $\ii 0^+$ prescription of the first line is natural: All causality flows naturally from left to right and results in a retarded worldline propagator.
Instead, in the second line, the causality flow does not follow the external lines.
Due to its symmetry, however, it must be symmetric in retarded and advanced propagators.
The powers of propagators on the right-hand-sides are then determined by power counting.
On top, a careful analysis shows that the (real) $\eps$-dependent factor of proportionality $f(\eps)$ is the same when an appropriate factor of $\ii$ is inserted in the first line.

At this point, having determined $f(\eps)$, the remaining integration for the four tail-of-tail boundaries is straightforward.
Namely, it is now a one-loop integral with generalized worldline propagators: $M[\omega_+^\alpha \omega_-^\beta]$.
The function $f(\eps)$ is determined from the three-loop integral,
\begin{align}
  f(\eps)=
  \int_{\boldsymbol{\ell}_1, \boldsymbol{\ell}_2, \boldsymbol{\ell}_3} 
&\frac{1}
{(\boldsymbol{\ell}_1 - \boldsymbol{\ell}_2)^2 (\boldsymbol{\ell}_2 - \boldsymbol{\ell}_3)^2 
(\boldsymbol{\ell}_1^2 +1)
 (\boldsymbol{\ell}_3^2 +1)} 
 \,,
\end{align}
with $(3-2\eps)$-dimensional Euclidean bold-face momenta.


\subsection{Brandenburg Gate integrals}
\begin{figure}[!h]
  \begin{tikzpicture}
  \coordinate (inA) at (0.2,.6);
    \coordinate (outA) at (5,.6);
    \coordinate (inB) at (0.2,-.6);
    \coordinate (outB) at (5,-.6);
    \coordinate (1A) at (.8,.6);
    \coordinate (2A) at (2,.6);
    \coordinate (3A) at (3.2,.6);
    \coordinate (4A) at (4.4,.6);
    \coordinate (1B) at (.8,-.6);
    \coordinate (2B) at (2,-.6);
    \coordinate (3B) at (3.2,-.6);
    \coordinate (4B) at (4.4,-.6);
    \coordinate (23C) at (2.6, .8);
    \draw [fill] (1A) circle (.08);
    \draw [fill] (2A) circle (.08);
    \draw [fill] (3A) circle (.08);
    \draw [fill] (4A) circle (.08);
    \draw [fill] (1B) circle (.08);
    \draw [fill] (2B) circle (.08);
    \draw [fill] (3B) circle (.08);
    \draw [fill] (4B) circle (.08);
    
     \draw [dotted] (inA) -- (outA);
     \draw [dotted] (inB) -- (outB);
     \draw [photon] (1A) -- (1B);
     \draw [photon] (2A) -- (2B);
     \draw [photon] (3A) -- (3B);
     \draw [photon] (4A) -- (4B);
     
     \draw [zParticle] (1A) -- (2A);
     \draw [photon] (2A) to[out=80, in=100] (3A);
     \draw [zParticle] (4A) -- (3A); 
     \draw [zParticle] (2,0.9) to[out=60, in=120] (3.2,0.9);

     \draw (1.4, 0.0) node [below] {$f(\tau_1)$};
     \draw (2.6, -0.7) node [below] {$G(\tau_1 - \tau_2)$};
     \draw (3.85, 0.0) node [below] {$f(\tau_2)$};

\end{tikzpicture}
\end{figure}

Consider the integral $M[\omega_+,g_+,\omega_-]$, nicknamed ``Brandenburg Gate" for its resemblance to the real-world structure. The only loop-by-loop integration we may perform here is the graviton bubble, leaving behind a genuine three-loop integral, for which the integral over Schwinger parameters is too difficult. However, since we only need the first two orders in $\epsilon$, it suffices to find identities at the level of the series expansion. Here, we show how such identities may be derived. In position-space, we can express it as an integral over the time-domain:
\be
A = M[\omega_+,g_+,\omega_-]\left(|b|, \epsilon \right) = \int \mathrm d\tau_1 \mathrm d\tau_2 \, f(\tau_1) G(\tau_1-\tau_2) f(\tau_2) \, ,
\ee
where $f(\tau)$ represents the loops to the side while $G(\tau_1-\tau_2)$ represents the middle integral including the bubble.
Because of $\tau_1  \leftrightarrow \tau_2$ symmetry of the integral, one may assume $G(\tau) = G(-\tau)$. Now consider the following two integrals. By partial fraction identities:
\be
\begin{aligned}
M[\omega_+, \omega_+, g_+] &= \frac1{2} M_1[\omega_+, \omega_+, g_+] \, , \\
M[\omega_-, \omega_-, g_+] &= \frac1{2}M_1[\omega_-, \omega_-, g_+] \, ,
\end{aligned}
\ee
such that
\be
M[\omega_+, \omega_+, g_+] \left(|b|, \epsilon \right) = \frac1{2} \int \mathrm d\tau_1 \mathrm d\tau_2 \, f(\tau_1) f(\tau_1) G(\tau_1-\tau_2) \, .
\ee	
We would like to put $M[\omega_-, \omega_-, g_+]$ in a similar form:
\be
\begin{aligned}
M[\omega_-, \omega_-, g_+] \left(|b|, \epsilon \right) &= \frac1{2} \int \mathrm d\tau_1 \mathrm d\tau_2 \, f(-\tau_1) f(-\tau_1) G(\tau_1-\tau_2) \\
&= \frac1{2} \int \mathrm d\tau_1 \mathrm d\tau_2 \, f(\tau_2) f(\tau_2) G(\tau_1-\tau_2) \, ,
\end{aligned}
\ee
where in the last step we have changed variables $\tau_1 \rightarrow -\tau_2$, $\tau_2 \rightarrow -\tau_1$. Then define the symmetric quantity:
\be
\begin{aligned}
B &= M[\omega_+, \omega_+, g_+]\left(|b|, \epsilon \right)  + M[\omega_-, \omega_-, g_+]\left(|b|, \epsilon \right)  \\
&= \frac1{2} \int \mathrm d\tau_1 \mathrm d\tau_2 \, \left(f^2(\tau_1) + f^2(\tau_2)\right) G(\tau_1-\tau_2) \, .
\end{aligned}
\ee
By computation, it is found that 
\be
\begin{aligned}
f(\tau) &= \frac{c_0}{\epsilon} + c_1(\tau) + \mathcal{O}(\epsilon)\, ,
\end{aligned}
\ee
and that $G(\tau)$ stars at order $\eps^1$.
Therefore, expanding $A$ and $B$ yields:
\be
\begin{aligned}
A &= \int \mathrm d\tau_1 \mathrm d\tau_2 \left(\frac{c_0^2}{\epsilon^2} + \frac{c_0}{\epsilon}\left(	c_1 (\tau_1) + c_1(\tau_2) \right) + \mathcal{O}(\epsilon)\right) G(\tau_1 - \tau_2) \\
&= \int \mathrm d\tau_1 \mathrm d\tau_2 \left( \frac{c_0^2}{\epsilon^2} + 2\frac{c_0}{\epsilon} c_1(\tau_1)  \right) G(\tau_1 - \tau_2) + \mathcal{O}(\epsilon) \, , \\
B &= \frac1{2} \int \mathrm d\tau_1 \mathrm d\tau_2 \left( \frac{c_0^2}{\epsilon^2} + 2\frac{c_0}{\epsilon} c_1(\tau_1) +  \frac{c_0^2}{\epsilon^2} + 2\frac{c_0}{\epsilon} c_1(\tau_2) + \mathcal{O}(\epsilon)\right) G(\tau_1 - \tau_2) \\
&= \int \mathrm d\tau_1 \mathrm d\tau_2 \left( \frac{c_0^2}{\epsilon^2} + 2\frac{c_0}{\epsilon} c_1(\tau_1) \right) G(\tau_1 - \tau_2) + \mathcal{O}(\epsilon)\, .
\end{aligned}
\ee
Moreover, in position space, we have $A=B+ \mathcal{O}(\epsilon)$. 
In this case, the Fourier transform between position space and momentum space introduces a factor $1/\epsilon$.
Thus, the final statement in momentum space is:
\be
M[\omega_+,g_+,\omega_-]
=
-M[\omega_+, \omega_+, g_+] 
-M[\omega_-, \omega_-, g_+] + \mathcal{O}(\epsilon^0) \, .
\ee
As similar analysis confirms an equivalent relation for the odd-in-$v^\mu$ Brandenburg Gate integrals:
\begin{align}
  M[\omega_+,g_+ \omega_+,\omega_-]
=
-M[\omega_+, \omega_+, g_+ \omega_+] 
-M[\omega_-, \omega_-, g_+ \omega_+] + \mathcal{O}(\epsilon^0) \, .
\end{align}
By supplementing the calculation of cuts with these identities, we can determine the series expansion of Brandenburg Gate integrals with all worldline orientations up to $\mathcal{O}(\epsilon^0)$.

\section{SUPPLEMENTARY EQUATIONS}

\subsection{Series expansions of boundary integrals}

Here we provide all boundary integrals used in our calculation as a series expansion in $\epsilon$. Each integral has been normalized by a factor of $\exp(4\gamma_E \epsilon)\pi^{-4 \epsilon}$ for brevity. Additionally, the integrals also depend trivially on $\gamma$: for each $g$, we pull out a factor of $(\gamma^2-1)^{\frac1{2}-\epsilon}$,
and for each $\omega$ a factor of $1/\sqrt{\gamma^2-1}$. In addition, we put $|q|=1$. We have computed:
\allowdisplaybreaks

\subsubsection*{\textbf{PPR Integrals}} 
 \textit{Even in $v^\mu$} 
 \begin{subequations}
 \begin{flalign} 
 & M[0,g_+,0]=-\frac{1}{65536 \pi ^2}-\frac{(3+26 \log (2)) \epsilon }{65536 \pi ^2}+\mathcal{O}\!\left(\epsilon ^2\right) & \\
 & M[g_+,0,0]=-\frac{1}{8192 \pi ^4}-\frac{(17+6 \log (2)) \epsilon }{8192 \pi ^4}+\mathcal{O}\!\left(\epsilon ^2\right) & \\
 & M[0,g_+ \omega_+,\omega_-]=-\frac{1}{4096 \pi ^4 \epsilon ^2}-\frac{6+3 \log (2)}{2048 \pi ^4 \epsilon }+\mathcal{O}\!\left(\epsilon ^0\right) & \\
 & M[0,g_+ \omega_+,\omega_+]=\frac{1}{4096 \pi ^4 \epsilon ^2}+\frac{6+3 \log (2)}{2048 \pi ^4 \epsilon }+\mathcal{O}\!\left(\epsilon ^0\right) & \\
 & M[g_+,\omega_-,\omega_-]=-\frac{3}{16384 \pi ^4 \epsilon ^2}-\frac{13+9 \log (2)}{8192 \pi ^4 \epsilon }+\mathcal{O}\!\left(\epsilon ^0\right) & \\
 & M[g_+,\omega_-,\omega_+]=\frac{5}{16384 \pi ^4 \epsilon ^2}+\frac{19+15 \log (2)}{8192 \pi ^4 \epsilon }+\mathcal{O}\!\left(\epsilon ^0\right) & \\
 & M[g_+,\omega_+,\omega_-]=\frac{1}{16384 \pi ^4 \epsilon ^2}+\frac{-1+3 \log (2)}{8192 \pi ^4 \epsilon }+\mathcal{O}\!\left(\epsilon ^0\right) & \\
 & M[g_+,\omega_+,\omega_+]=\frac{1}{16384 \pi ^4 \epsilon ^2}+\frac{7+3 \log (2)}{8192 \pi ^4 \epsilon }+\mathcal{O}\!\left(\epsilon ^0\right) & \\
 & M[\omega_+,g_+,\omega_-]=\frac{1}{8192 \pi ^4 \epsilon ^2}+\frac{3+3 \log (2)}{4096 \pi ^4 \epsilon }+\mathcal{O}\!\left(\epsilon ^0\right) & \\
 & M[\omega_+,g_+,\omega_+]=\frac{1}{8192 \pi ^4 \epsilon ^2}+\frac{7+3 \log (2)}{4096 \pi ^4 \epsilon }+\mathcal{O}\!\left(\epsilon ^0\right) & \\
 & M[g_+ \omega_+,0,\omega_-]=-\frac{1}{4096 \pi ^4 \epsilon ^2}-\frac{6+3 \log (2)}{2048 \pi ^4 \epsilon }+\mathcal{O}\!\left(\epsilon ^0\right) & \\
 & M[g_+ \omega_+,0,\omega_+]=\frac{1}{4096 \pi ^4 \epsilon ^2}+\frac{6+3 \log (2)}{2048 \pi ^4 \epsilon }+\mathcal{O}\!\left(\epsilon ^0\right) & \\
 & M[g_+ \omega_+,\omega_-,0]=\frac{1}{8192 \pi ^2 \epsilon }+\mathcal{O}\!\left(\epsilon ^0\right) & \\
 & M[g_+ \omega_+,\omega_+,0]=-\frac{1}{8192 \pi ^2 \epsilon }+\mathcal{O}\!\left(\epsilon ^0\right) & 
 \end{flalign} 
\end{subequations}
 
 \textit{Odd in $v^\mu$} 
 \begin{subequations}
 \begin{flalign} 
 & M[0,g_+,\omega_-]=\frac{i}{768 \pi ^3}+\frac{55 i \epsilon }{2304 \pi ^3}+\mathcal{O}\!\left(\epsilon ^2\right) & \\
 & M[0,g_+,\omega_+]=\frac{i (-1+4 \log (2)) \epsilon }{768 \pi ^3}+\mathcal{O}\!\left(\epsilon ^2\right) & \\
 & M[0,g_+ \omega_+,0]=\frac{i}{1024 \pi ^3}+\frac{i (4+2 \log (2)) \epsilon }{256 \pi ^3}+\mathcal{O}\!\left(\epsilon ^2\right) & \\
 & M[g_+,0,\omega_-]=\frac{i}{1024 \pi ^3}+\frac{i (25+2 \log (2)) \epsilon }{1536 \pi ^3}+\mathcal{O}\!\left(\epsilon ^2\right) & \\
 & M[g_+,0,\omega_+]=-\frac{i}{3072 \pi ^3}+\frac{i (-41+30 \log (2)) \epsilon }{4608 \pi ^3}+\mathcal{O}\!\left(\epsilon ^2\right) & \\
 & M[g_+,\omega_-,0]=\frac{i}{2048 \pi ^3}+\frac{i (2+3 \log (2)) \epsilon }{512 \pi ^3}+\mathcal{O}\!\left(\epsilon ^2\right) & \\
 & M[g_+,\omega_+,0]=\frac{i}{2048 \pi ^3}+\frac{i (4+3 \log (2)) \epsilon }{512 \pi ^3}+\mathcal{O}\!\left(\epsilon ^2\right) & \\
 & M[g_+ \omega_+,0,0]=\frac{i}{1024 \pi ^3}+\frac{i (7+6 \log (2)) \epsilon }{512 \pi ^3}+\frac{i \left(272+9 \pi ^2+48 \log (2) (7+3 \log (2))\right) \epsilon ^2}{2048 \pi ^3}+\mathcal{O}\!\left(\epsilon ^3\right) & \\
 & M[\omega_+,g_+ \omega_+,\omega_-]=-\frac{i}{2048 \pi ^3 \epsilon ^2}-\frac{i (1+2 \log (2))}{1024 \pi ^3 \epsilon }+\mathcal{O}\!\left(\epsilon ^0\right) & \\
 & M[\omega_+,g_+ \omega_+,\omega_+]=\frac{i}{2048 \pi ^3 \epsilon ^2}+\frac{i (1+2 \log (2))}{1024 \pi ^3 \epsilon }+\mathcal{O}\!\left(\epsilon ^0\right) & \\
 & M[g_+ \omega_+,\omega_-,\omega_-]=\frac{i}{4096 \pi ^3 \epsilon ^2}+\frac{i (1+2 \log (2))}{2048 \pi ^3 \epsilon }+\mathcal{O}\!\left(\epsilon ^0\right) & \\
 & M[g_+ \omega_+,\omega_-,\omega_+]=-\frac{3 i}{4096 \pi ^3 \epsilon ^2}-\frac{3 i (1+2 \log (2))}{2048 \pi ^3 \epsilon }+\mathcal{O}\!\left(\epsilon ^0\right) & \\
 & M[g_+ \omega_+,\omega_+,\omega_-]=-\frac{3 i}{4096 \pi ^3 \epsilon ^2}-\frac{3 i (1+2 \log (2))}{2048 \pi ^3 \epsilon }+\mathcal{O}\!\left(\epsilon ^0\right) & \\
 & M[g_+ \omega_+,\omega_+,\omega_+]=\frac{i}{4096 \pi ^3 \epsilon ^2}+\frac{i (1+2 \log (2))}{2048 \pi ^3 \epsilon }+\mathcal{O}\!\left(\epsilon ^0\right) & 
 \end{flalign} 
\end{subequations}
 \subsubsection*{\textbf{PRR Integrals}} 
 \textit{Even in $v^\mu$} 
  \begin{subequations}
 \begin{flalign}
 & M[g_- g_+,0]=\frac{1}{30720 \pi ^4 \epsilon }+\frac{58+15 \log (2)}{115200 \pi ^4}+\mathcal{O}\!\left(\epsilon ^1\right) & \\
 & M[g_+ g_+,0]=-\frac{1}{30720 \pi ^4 \epsilon }-\frac{58+15 \log (2)}{115200 \pi ^4}+\mathcal{O}\!\left(\epsilon ^1\right) & \\
 & M[g_- g_+ \omega_+,\omega_+]=-\frac{1}{4096 \pi ^4 \epsilon }-\frac{7+2 \log (2)}{2048 \pi ^4}+\mathcal{O}\!\left(\epsilon ^1\right) & \\
 & M[g_+ g_+ \omega_+,\omega_-]=\frac{3}{4096 \pi ^4 \epsilon }+\frac{17+6 \log (2)}{2048 \pi ^4}+\mathcal{O}\!\left(\epsilon ^1\right) & \\
 & M[g_+ g_+ \omega_+,\omega_+]=-\frac{1}{4096 \pi ^4 \epsilon }-\frac{3+2 \log (2)}{2048 \pi ^4}+\mathcal{O}\!\left(\epsilon ^1\right) &
 \end{flalign} 
 \end{subequations}
 
 \textit{Odd in $v^\mu$} 
 \begin{subequations}
 \begin{flalign} 
 & M[g_- g_+,\omega_+]=\frac{i}{8192 \pi ^3 \epsilon }+\frac{i (7+10 \log (2))}{8192 \pi ^3}+\mathcal{O}\!\left(\epsilon ^1\right) & \\
 & M[g_+ g_+,\omega_-]=\frac{i}{8192 \pi ^3 \epsilon }+\frac{i (7+10 \log (2))}{8192 \pi ^3}+\frac{i \left(342+5 \pi ^2+60 \log (2) (7+5 \log (2))\right) \epsilon }{49152 \pi ^3}+\mathcal{O}\!\left(\epsilon ^2\right) & \\
 & M[g_+ g_+,\omega_+]=-\frac{i}{8192 \pi ^3 \epsilon }-\frac{i (7+10 \log (2))}{8192 \pi ^3}-\frac{i \left(342+29 \pi ^2+60 \log (2) (7+5 \log (2))\right) \epsilon }{49152 \pi ^3}+\mathcal{O}\!\left(\epsilon ^2\right) & \\
 & M[g_- g_+ \omega_+,0]=0 & \\
 & M[g_+ g_+ \omega_+,0]=-\frac{i \epsilon }{384 \pi ^3}+\frac{i (-34+9 \log (2)) \epsilon ^2}{576 \pi ^3}+\mathcal{O}\!\left(\epsilon ^3\right) & 
 \end{flalign} 
 \end{subequations}
 \subsubsection*{\textbf{RRR Integrals}} 
 \textit{Even in $v^\mu$} 
 \begin{subequations}
 \begin{flalign} 
 & M[g_- g_+ g_+]=-\frac{1}{12288 \pi ^4}-\frac{(41+6 \log (2)) \epsilon }{36864 \pi ^4}+\mathcal{O}\!\left(\epsilon ^2\right) & \\
 & M[g_+ g_+ g_+]=\frac{1}{4096 \pi ^4}+\frac{(41+6 \log (2)) \epsilon }{12288 \pi ^4}+\frac{\left(587-6 \pi ^2+3 \log (2) (41+3 \log (2))\right) \epsilon ^2}{18432 \pi ^4}+\mathcal{O}\!\left(\epsilon ^3\right) & \\
 & M[\text{tail}_+ \text{tail}_+]=-\frac{3}{4096 \pi ^4 \epsilon }-\frac{3 (8+\log (2))}{2048 \pi ^4}-\frac{3 (90+\log (2) (16+\log (2))) \epsilon }{2048 \pi ^4}+\mathcal{O}\!\left(\epsilon ^2\right) & \\
 & M[\text{tail}_+ \text{tail}_-]=-\frac{1}{4096 \pi ^4 \epsilon }-\frac{8+\log (2)}{2048 \pi ^4}+\mathcal{O}\!\left(\epsilon ^1\right) &
 \end{flalign} 
 \end{subequations}
 
 \textit{Odd in $v^\mu$} 
 \begin{subequations}
 \begin{flalign} 
 & M[g_- g_+ g_+ \omega_+]=\frac{i}{4096 \pi ^3}+\frac{i (1+4 \log (2)) \epsilon }{1024 \pi ^3}+\mathcal{O}\!\left(\epsilon ^2\right) & \\
 & M[g_+ g_+ g_+ \omega_+]=-\frac{i}{4096 \pi ^3}-\frac{i (1+4 \log (2)) \epsilon }{1024 \pi ^3}+\mathcal{O}\!\left(\epsilon ^2\right) & \\
 & M[\text{tail}_+ \text{tail}_+ \omega_+]=\frac{i}{1024 \pi ^3 \epsilon }+\frac{i (1+2 \log (2))}{128 \pi ^3}+\mathcal{O}\!\left(\epsilon ^1\right) & \\
 & M[\text{tail}_+ \text{tail}_- \omega_+]=0 &
 \end{flalign}
 \end{subequations}
 The PPP boundary integrals can be decomposed into simple products of $\Gamma$-functions that can be obtained from the 
 loop-by-loop methods given above. 
 
\subsection{Additional $G_i$ functions}
We list here all 20 new transcendental functions necessary to derive the canonical form of the differential equations:
\begin{subequations}\label{eq:GFunctions}
\begin{align}
      G_1'(x)&=-\frac{96 x \left(x^4+1\right) \varpi_{0}(x)^2}{(x-1)^2 (x+1)^2 \left(x^2+1\right)^2 \alpha_1(x)} \, , \\
      G_2'(x)&=-\frac{16 \left(7 x^{12}+314 x^{10}+329 x^8+1340 x^6+329 x^4+314 x^2+7\right) \varpi_{0}(x)^2}{3 (x-1)^3 x (x+1)^3 \left(x^2+1\right)^3} \\\nonumber
      	&+\frac{16 \left(7 x^8+136 x^6+42 x^4+136 x^2+7\right) \varpi_{0}(x)^2 \alpha_1 '(x)}{3 (x-1)^2 (x+1)^2 \left(x^2+1\right)^2 \alpha_1(x)}-\frac{2 \left(5 x^8+28 x^6+262 x^4+28 x^2+5\right) \varpi_{0}(x)^2 \alpha_1 '(x)^2}{3 (x-1) x (x+1) \left(x^2+1\right) \alpha_1(x)^2} \\\nonumber
	&-\frac{x G_1(x)^2 \alpha_1(x)^2}{(x-1) (x+1) \left(x^2+1\right) \varpi_{0}(x)^2} \, ,\\
      G_3'(x)&=-\frac{x G_1(x) \alpha_1(x)^2}{(x-1) (x+1) \left(x^2+1\right) \varpi_{0}(x)^2} \, ,\\
      G_4'(x)&= -\frac{16 \left(7x^8+136 x^6+42 x^4+136 x^2+7\right) \varpi_{0}(x)^2}{3 (x-1)^2 x (x+1)^2 \left(x^2+1\right)^2 \alpha_1(x)} +\frac{4 \left(5 x^8+28 x^6+262 x^4+28 x^2+5\right) \varpi_{0}(x)^2 \alpha_1 '(x)}{3 (x-1) x^2 (x+1) \left(x^2+1\right) \alpha_1(x)^2} \\\nonumber
      	&+\frac{G_2(x)}{x\alpha_1(x)} \, ,\\
      G_5'(x)&=\varK(x) \varpi_{0}''(x) \, , \\
      G_6'(x)&=\varK(x) \varpi_{0}'(x) \, , \\
      G_7'(x)&=\frac{\varK(x) G_3(x) \varpi_{0}'(x)}{\alpha_1(x)} \, , \\
      G_8'(x)&=\frac{\varK(x) G_3(x) \varpi_{0}(x) \alpha_1 '(x)}{\alpha_1(x)^2} \, , \\
      G_9'(x)&=\frac{\varK(x) \varpi_{0}'(x)}{\alpha_1(x)}-\frac{\varK(x) \varpi_{0}(x) \alpha_1 '(x)}{2 \alpha_1(x)^2} \, ,\\
      G_{10}'(x)&=-\frac{\left(3 x^4-2 x^2+3\right)\varK(x) \varpi_{0}(x)}{32 x^2}+\frac{4 x^3 \left(3 x^2-5\right)G_5(x)}{32 x^2}+\frac{(27 x^4-10 x^2+3)G_6(x)}{32 x^2} \, ,\\
      G_{11}'(x)&=-\frac{3 (x-1) (x+1) \left(x^2+1\right)\varK(x) \varpi_{0}(x)}{8 x^2}+\frac{12 x^5G_5(x)}{8 x^2}   +\frac{(27 x^4-4 x^2-3)G_6(x)}{8 x^2} \, ,\\
	G_{12}'(x)&=-\frac{(x^{12}+2 x^{10}-73 x^8+236 x^6-73 x^4+2 x^2+1)\varK(x)\varpi_{0}(x)}{16 (x-1)^2 x^2 (x+1)^2 \left(x^2+1\right)^2}-\frac{\left(x^2-3\right) \left(3 x^2-1\right)\varK(x)G_1(x)\alpha_1(x)}{32 (x-1) (x+1) \left(x^2+1\right)\varpi_{0}(x)} 
  \nonumber\\\nonumber
	&+\left[  \frac{3 (x-1) (x+1) \left(x^2+1\right)\varK(x)\varpi_{0}(x)}{16 x^2\alpha_1(x)} -\frac{\left(x^2-3\right) \left(3 x^2-1\right)G_5(x)}{32 x\alpha_1(x)} -\frac{(9 x^4-10 x^2-3)G_6(x)}{32 x^2\alpha_1(x)}     \right] G_3(x) 
  \\\nonumber
	&+\frac{\left(x^2-3\right) \left(3 x^2-1\right) \left(x^4-16 x^2+1\right)G_5(x)}{48 (x-1) x (x+1) \left(x^2+1\right)}+ \frac{(9 x^{12}-143 x^{10}-85 x^8+624 x^6-551 x^4+47 x^2+3)G_6(x)}{24 (x-1)^2 x^2 (x+1)^2 \left(x^2+1\right)^2} 
  \\
	&-\frac{(9 x^4-10 x^2-3)G_7(x)}{16 x^2} + \frac{(9 x^4-10 x^2-3)G_8(x)}{32 x^2}-\frac{2 \left(x^4-16 x^2+1\right)G_{10}(x)}{3 (x-1) x (x+1) \left(x^2+1\right)} \, \\
	G_{13}'(x)&= -\frac{15 \left(x^2+1\right)\varpi_{0}(x)}{(x-1)^2 (x+1)^2}-\frac1{\varK(x)} \left[    \frac{13 \left(x^2-3\right) \left(3 x^2-1\right)G_5(x)}{6 (x-1) x (x+1)} +\frac{13 \left(9 x^4-10 x^2-3\right)G_6(x)}{6 (x-1) x^2 (x+1)} \right. \\\nonumber
		&\left.-\frac{208G_{10}(x)}{3 (x-1) x (x+1)}    \right] \, , \\
%
%
	G_{14}'(x)&=-\frac1{\varK(x)} \left[  -\frac{3 \left(x^2+1\right)G_5(x)}{4 x}  -\frac{(9 x^4-4 x^2+3)G_6(x)}{4 (x-1) x^2 (x+1)} +\frac{2G_{11}(x)}{(x-1) x (x+1)}         \right] \, , \\
%
%
	G_{15}'(x)&=   -\frac{(x^8-12 x^6-2 x^4-12 x^2+1)\varK(x) \varpi_{0}(x)}{4 (x-1) x^2 (x+1) \left(x^2+1\right)} - \frac {3\varK(x)G_1(x)\alpha_1(x)}{8\varpi_{0}(x)} \\\nonumber
		&+ \left[   \frac{(3 x^4-2 x^2+3)\varK(x) \varpi_{0}(x)}{4 x^2\alpha_1(x)} -\frac{3 (x-1) (x+1) \left(x^2+1\right)G_5(x)}{8 x\alpha_1(x)} -\frac{(9 x^4-4 x^2+3)G_6(x)}{8 x^2\alpha_1(x)} \right. \\\nonumber
		&\left. +\frac{G_{11}(x)}{x\alpha_1(x)}  \right]G_3(x) + \frac{(x^4-16 x^2+1)G_5(x)}{4 x} + \frac{(9 x^8-89 x^6+32 x^4-11 x^2+3)G_6(x)}{6 (x-1) x^2 (x+1) \left(x^2+1\right)}-\frac{3 \left(3 x^4+1\right)G_7(x)}{4 x^2} \\\nonumber
		&+\frac{3 \left(3 x^4+1\right)G_8(x)}{8 x^2}-\frac{2 \left(x^4-16 x^2+1\right)G_{11}(x)}{3 (x-1) x (x+1) \left(x^2+1\right)} \, , \\  
%
%
	G_{16}'(x)&=-\frac{3 (x-1) (x+1) \left(x^2+1\right)\varK(x) \varpi_{0}(x)}{16 x^2\alpha_1(x)} + \frac{\left(x^2-3\right) \left(3 x^2-1\right)G_5(x)}{32 x\alpha_1(x)}+\frac{(9 x^4-10 x^2-3)G_6(x)}{32 x^2\alpha_1(x)} \\\nonumber
		&+\frac{(9 x^4-10 x^2-3)G_9(x)}{16 x^2}  - \frac{G_{10}(x)}{x\alpha_1(x)} \, ,\\
      G_{17}'(x)&=-\frac{\left(3 x^4-2 x^2+3\right)\varK(x) \varpi_{0}(x)}{4 x^2 \alpha_1(x)} -\frac{3 (x-1) (x+1) \left(x^2+1\right)G_5(x)}{8 x\alpha_1(x)} + \frac{(9 x^4-4 x^2+3)G_6(x)}{8 x^2\alpha_1(x)} \\\nonumber
      	& + \frac{3 \left(3 x^4+1\right)G_9(x)}{4 x^2} - \frac{G_{11}(x)}{x\alpha_1(x)} \, , \\ 
      G_{18}'(x)&=\frac{\varpi_{0}(x)}{(x-1)^2} \, ,\\
      G_{19}'(x)&=\frac{\varpi_{0}(x)}{x} \, ,\\
      G_{20}'(x)&=\frac{\varpi_{0}(x)}{(x+1)^2} \, ,
\end{align}
\end{subequations}
where $\varK=(\frac{2}{\pi})^2 K^2(1-x^2)$ and $f'(x)=\dfrac{\mathrm d }{\mathrm dx}f(x)$.

\end{widetext}

\bibliographystyle{naturemag}
\bibliography{5pm-dissipative}

\end{document}